\documentclass[letter,12pt]{article}
\usepackage{graphicx,amssymb,amsmath,color}

\def\beq{\begin{equation}}
\def\eeq{\end{equation}}

\def\bea{\begin{eqnarray}}
\def\eea{\end{eqnarray}}
\def\bmat{\begin{pmatrix}}
\def\emat{\end{pmatrix}}
\def\bei{\begin{itemize}}
\def\eei{\end{itemize}}
\def\xfb{\, {\rm fb}}
\def\={\,=\,}
\def\+{\,+\,}
\def\-{\,-\,}

\def\GeV{\, {\rm GeV}}

\def\TeV{\, {\rm TeV}}

\def\fb{\, {\rm fb}}

\newcommand{\Fig}[1]{Fig.~\ref{#1}}

\newcommand{\Eq}[1]{eq.(\ref{#1})}

\makeatletter
\renewcommand{\section}{\@startsection{section}{1}{0em}%
        {-3.25ex \@plus -1ex \@minus -.2ex}%
        {2.0ex \@plus.2ex}%
        {\normalfont\large\bfseries}}
\renewcommand{\subsection}{\@startsection{subsection}{2}{0em}%
        {-2.75ex\@plus -1ex \@minus -.2ex}%
        {1.25ex \@plus .2ex}%
        {\normalfont\bfseries}}
\renewcommand{\subsubsection}%
        {\@startsection{subsubsection}{3}{0em}%
        {-2.0ex\@plus -1ex \@minus -.2ex}%
        {1.0ex \@plus .2ex}%
        {\normalfont\itshape}}
\makeatother

\def\To{\Rightarrow}

\def\tT{t\bar{t}}

\def\KKg{g^{(1)}}

\begin{document}
\baselineskip=17pt

\begin{titlepage}

\noindent
\begin{flushright}
CERN-PH-TH/2010-170\\
MCTP-10-26 \\

\end{flushright}
\vspace{1cm}

\begin{center}
  \begin{Large}
    \begin{bf}
Low-scale warped extra dimension \\
and its predilection for multiple top quarks
        \end{bf}
  \end{Large}
\end{center}

\vspace{0.5cm}
\begin{center}
\begin{Large}
Sunghoon Jung, James D. Wells \\
\end{Large}

\vspace{0.3cm}
\begin{it}
CERN, Theory Group, CH-1211 Geneva 23, Switzerland, {\rm and} \\
Physics Department, University of Michigan, Ann Arbor, MI 48109, USA \\
\vspace{0.1cm}

\vspace{0.1cm}
\end{it}

\vspace{0.5cm}
\end{center}

\begin{abstract}

Within warped extra dimension models that explain flavor through geometry, flavor changing neutral current constraints generally force the Kaluza-Klein scale to be above many TeV. This creates tension with a natural electroweak scale. On the other hand, a much lower scale compatible with precision electroweak and flavor changing neutral current constraints is allowed if we decouple the Kaluza-Klein states of Standard Model gauge bosons from light fermions   ($c_{\rm light}\simeq c_b\simeq 0.5$ bulk mass parameters). The main signature for this approach is four top quark production via the Kaluza-Klein excitations' strong coupling to top quarks. We study single lepton, like-sign dilepton, and trilepton observables of four-top events at the Large Hadron Collider. The like-sign dilepton signature typically has the largest discovery potential for a strongly coupled right-handed top case ($M_{KK} \sim 2-2.5 \TeV$), while single lepton is the better when the left-handed top couples most strongly ($M_{KK} \sim 2 \TeV$). We also describe challenging lepton-jet collimation issues in the like-sign dilepton and trilepton channels. An alternative single lepton observable is considered which takes advantage of the  many bottom quarks in the final state. Although searches of other particles may compete, we find that four top production via Kaluza-Klein gluons is most promising in a large region of this parameter space.

\end{abstract}

\vspace{1cm}

\begin{flushleft}
\begin{small}
August 2010
\end{small}
\end{flushleft}

\end{titlepage}

\tableofcontents

\section{Introduction}

The origin of heaviness of the third generation particles and its possible connection to physics responsible for electroweak symmetry breaking (EWSB) are still mysterious. Nevertheless, many candidates of new physics generically end up with either preferential couplings to the third generation or light partners of third generation. For example, top quark condensation and its mass are directly tied to electroweak scale dynamics in top-color models, and top partners are present and possibly lighter than other partners in the supersymmetry and Little Higgs models.

An extreme case is when new physics couples only to the third generation. This raises a challenge in discovering such physics at the collider because they are not produced directly from colliding partons/leptons. A gluon, not being a fermion, is an exception to this discussion, and can couple to colored new particles. However, such interaction vertices involve two or more new particles at leading order, which then are only produced in pairs in simple processes.

Meanwhile, a warped extra dimensional Randall-Sundrum (RS) model \cite{Randall:1999ee,Randall:1999vf} is a promising theory that attempts to explain the large hierarchy between Planck scale and weak scale. Exponentially warped background geometry is responsible for the huge difference of mass scales between two 4D spaces of moderate distance along the extra dimension.

The warped space in the RS model has been further feted by its ability to generate the flavor hierarchy \cite{Gherghetta:2000qt,Grossman:1999ra}.  The Higgs boson, being localized on the IR brane, feels only the exponentially warped tail of the bulk wave functions of UV localized fields. By properly localizing fermions, a wide range for the fermion mass spectrum can be obtained with anarchic Yukawa couplings. However, this flavor dependence inevitably induces flavor changing neutral currents (FCNC) mediated by Kaluza-Klein (KK) gauge bosons, and the RS model of this type generically in conflict with precise measurements of flavor physics. The strongest among them, for example, are from CP violation $\epsilon_K$ of the $K -\bar{K}$ system which requires the mass of KK states to be $M_{KK} \gtrsim 20 \TeV$ \cite{Csaki:2008zd,Blanke:2009}. When the Higgs is in the bulk, bounds from $\epsilon_K$ and $\epsilon^\prime / \epsilon_K$ of the $K^0 \to 2\pi$ process can be relieved to be $M_{KK} \gtrsim 5.5 \TeV$ \cite{Agashe:2008uz,Gedalia:2009ws} which is still well above the electroweak scale and beyond the collider reach if the anarchic Yukawa approach is pursued.

On the other hand, if one's highest priority is first and foremost to explain the Planck-weak hierarchy, the RS model can be made much more compatible with flavor measurements by assuming $c_{light}=0.5$\footnote{Our definition of the bulk mass parameter $c$ is the usual one, and matches, for example, that given in ref. \cite{Gherghetta:2000qt}. By $c_{light}$, we mean bulk masses of both left-handed (LH) and right-handed (RH) first two generations.} which decouples KK gauge bosons from Standard Model (SM) fermions and by having some flavor structure. Several flavor structures of the quark \cite{Fitzpatrick:2007sa,Csaki:2009wc,Cacciapaglia:2007fw} and lepton \cite{history:flavor-l,Chen:2008qg} sectors have been discussed in the literature. Among them, flavor universality in the RH down sector is very useful to make the theory consistent with $\epsilon_K$ measurements with mildly heavy $M_{KK} \gtrsim 4 \TeV$ \cite{Santiago:2008vq}. This is made possible by getting rid of the chiral enhanced left-right mixed current contributions to $\epsilon_K$ \cite{Csaki:2008zd}. On the other hand, in the geometric approach to the flavor hierarchy, there still exists some tension with several flavor observables that hover around the current bounds.

Another class of flavor structure is to align bulk masses with proper combinations of anarchic Yukawa couplings \cite{Fitzpatrick:2007sa}. The desired degree of alignment in the down-sector can be achieved by some bulk flavor symmetries \cite{Csaki:2009wc}, and the flavor bound is as low as the bound obtained from generic electroweak precision test (EWPT) $M_{KK} \gtrsim 3 \TeV$ \cite{Agashe:2003zs}. Furthermore, if the full flavor symmetry $SU(3)^5$ is gauged in the bulk and if (fully) \emph{hierarchical} Yukawa couplings generated from the flavor breaking at the UV brane are shined to the IR brane, minimal-flavor-violation is generically obtained and fermion localization is released from the duty of generating the hierarchy \cite{Delaunay:2010dw,Rattazzi:2000hs}. Consequently, universal $c_{light} \simeq c_b \simeq 0.5$ is a preferred solution for lighter KK states $M_{KK} \sim 1.5-2 \TeV$ compatible with flavor physics as well as with EWPT \cite{Delaunay:2010dw}.

Model independent global fit studies of EWPT in refs.~\cite{Carena:2006bn,Carena:2007ua} have also shown that the parameter space $c_{light} \sim 0.5$ is a consistent solution allowing light KK states $ M_{KK} \sim 2-3\TeV$ in RS models with custodial symmetry \cite{Agashe:2003zs,Agashe:2006at}. Very importantly, decoupling of SM KK gauge bosons from SM fermions can render the $S$ parameter small \cite{Agashe:2003zs}. $c_b$, the bulk mass of RH bottom quark, is also somewhat consistently allowed to be around $c_b \sim 0.5$. Although the absolute minimum of the fit may be obtained with $c_{light}$ and $c_b$ slightly away from $0.5$, this slight deviation is inconsequential if KK states are slightly heavier.

We also remark that by utilizing this special property of the $c_{light} \sim 0.5$ region with regard to EWPT, warped models of Higgless theory are made more successful \cite{Cacciapaglia:2004rb,Foadi:2004ps}. In this type of theory, light KK states are essential to recover  unitarity around the TeV scale.

Whatever the underlying reason may be for it, if $c_{light}=0.5$ is realized in nature, it could be difficult to discover RS physics in the standard channel $q\bar{q} \to \KKg \to \tT$. In this paper, we study alternative collider signatures of the RS model with $c_{light}=0.5$. {\it Four} top quarks can be abundantly produced via
\begin{itemize}
  \item $q\bar{q},gg \, \to \, \tT \KKg$ associate production
  \item $gg \, \to \, \KKg \KKg$ pair production
\end{itemize}
followed by $\KKg \to \tT$ decays. Four top quarks can then give rise to exciting collider signatures involving many leptons, bottom quarks as well as sizable missing energy. We categorize final states by the number of leptons (by ``leptons" we mean electrons or muons here):
\begin{itemize}
  \item single lepton (of any charge)
  \item like-sign dilepton (two and only two leptons of the same charge)
  \item trilepton (three leptons of any charges).
\end{itemize}
We will aim to estimate the discovery reach of the above multi-lepton channels, but not to reconstruct four top events nor a heavy resonance. That is for a later study. Also, regarding the single lepton category, in the case of nonzero left-handed top/bottom coupling, $\tT b\bar{b}$ events can also contribute to single lepton final states. This process will be considered as well.

In this paper, we will somewhat model independently assume the parameter space of $c_{light}=0.5$ for the first two generations to decouple KK-gluon from SM light fermions, and assume the flavor universal structure at least in the RH down-sector, i.e. $c_b = 0.5$ universal to $c_{light}$. As discussed, this parameter space can approximately represent the attractive solution of light $M_{KK} \sim 1.5 -3 \TeV$ obtained from the EWPT point of view \cite{Agashe:2003zs,Carena:2006bn,Carena:2007ua} as well as from the shining model of ref.\cite{Delaunay:2010dw}. Relevant theoretical issues with such parameter choices will be reviewed and discussed in section~\ref{sec:theory}.

Four top events have been discussed in several different contexts. Associate production ($t\bar tg^{(1)}$) in RS models have been studied without emphasis on $c_{light} \simeq 0.5$ in ref.\cite{Guchait:2007jd,Djouadi:2007eg}. These references used kinematic cuts on top quarks themselves which do not take into account topological characteristics of events such as overlap of objects, missing energy, number of objects, etc. Compositeness of the top quark can also be probed in the four top events \cite{Lillie:2007hd,Cheung:1995eq,Spira:1997ce,Pomarol:2008bh}; results with the like-sign dilepton (LSDL) observable in ref.\cite{Lillie:2007hd} agree with ours. Pair of gluinos in supersymmetry models \cite{Acharya:2009gb,Toharia:2005gm}, exotic fermions mixing with top quark strongly \cite{Contino:2008hi}, and $Z^\prime$ coupling preferentially to top quark \cite{Brooijmans:2010tn} can also give rise to four top events.

We first set the model framework and input parameters in section 2. Some theoretical thoughts on issues regarding $c_{light}=0.5\,(=c_b)$ will be reviewed. Then we present our Monte Carlo simulation results of discovery potential in sections 3 to 6. One can find the discussions on boosted leptonic top in section \ref{sec-leptop}. Section 7 is devoted to discussing other possible collider signatures competing with four top events and sorting out parameter space where four top events are most important.

\section{Model setup}

\subsection{Randall-Sundrum model with custodial symmetries}

The Randall-Sundrum model is a five-dimensional theory in the AdS background geometry~\cite{Randall:1999ee}. The metric is given by
\beq
ds^2 \= e^{-2ky} \eta_{\mu\nu} dx^\mu dx^\nu \, - \, dy^2
\eeq
where $x^\mu\, (\mu = 0,1,2,3)$ are 4D coordinates, and $y$ is the coordinate of the extra dimension. The extra dimension is compactified to the finite interval $0 \le y \le \pi r_c$. The curvature scale $k$ is assumed to be $k = M_{pl}$ in numerical computation. Exponential warping generates weak scale $M_{IR}$ from the Planck scale
\beq
\frac{M_{IR}}{k} \= e^{-\pi k r_c} = 10^{-16}, \qquad \qquad M_{IR} \sim {\cal O}(1 \TeV).
\label{eq:ir-k}\eeq
The Higgs boson is assumed to be localized on the IR brane, but the main discussion in this paper may be applicable to bulk Higgs case as well.

We assume that electoweak gauge symmetry in the bulk is enhanced with a custodial symmetry to be $SU(2)_L \times SU(2)_R \times U(1)_X$ \cite{Agashe:2003zs} with a discrete parity $P_{LR}$ exchanging $SU(2)_L$ and $SU(2)_R$ \cite{Agashe:2006at}. Custodial symmetry is essential to be consistent with electroweak precision tests  (EWPT), and to protect the accurately measured $Z b \bar{b}$ coupling \cite{Agashe:2006at,Carena:2006bn,Cacciapaglia:2006gp,Casagrande:2010si}. We emphasize that since Higgs-gauge contributions to the $T$ parameter has to be suppressed, custodial symmetry is required regardless of fermion couplings \cite{Agashe:2003zs}. There are extra KK particles in order to embed SM particles into extended symmetry representation. These extra particles will not be needed for our main discussion on four top events, but they may contribute to additional collider signatures. More discussion on these aspects can be found in section \ref{sec:other-sig}.

\begin{figure}
\centering
\includegraphics[angle=0,width=0.46\textwidth]{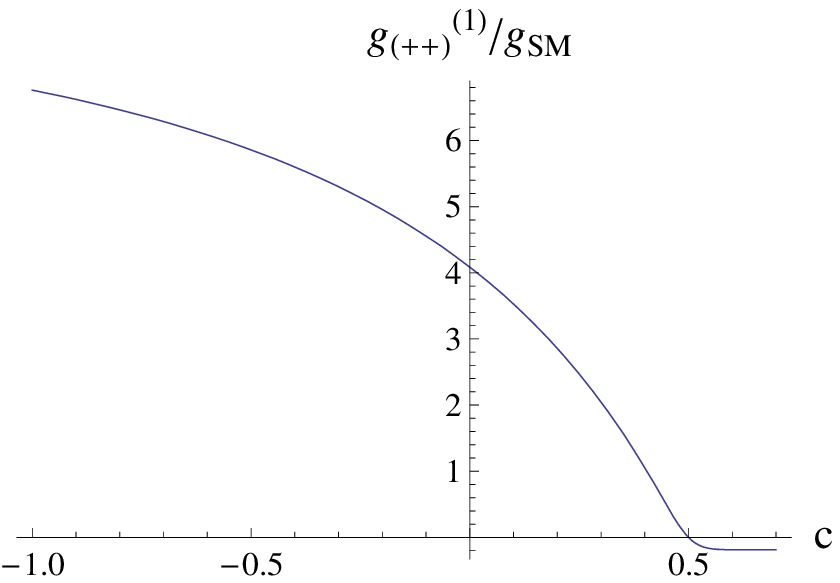}
\includegraphics[angle=0,width=0.46\textwidth]{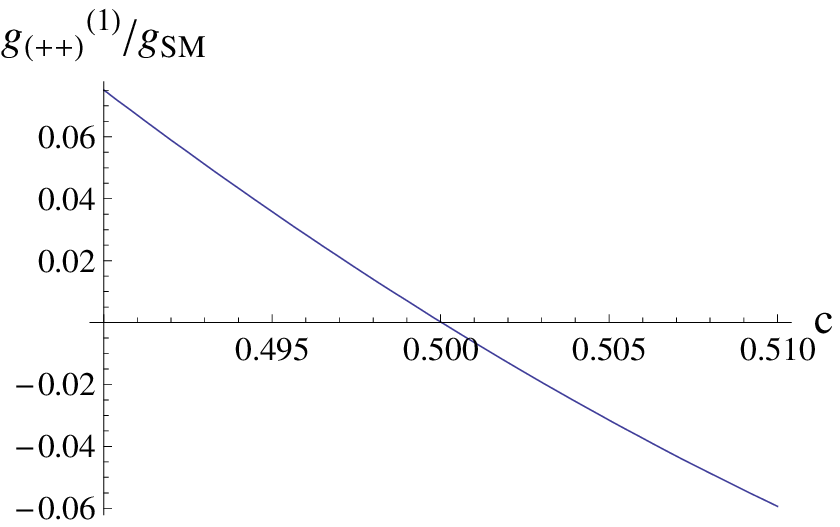}
\caption{Gauge coupling of zero mode fermions with KK gluon (the first KK gauge boson with $(+,+)$ boundary condition). $g_{KK}=0$ for $c=0.5$, and reaches its asymptotic value $g_{KK} \simeq -0.2$ for higher $c$.}
\label{coup-f0f0g1} \end{figure}

Our focus will be on the KK excitation of the gluon $\KKg$ because this feels strong coupling. $\KKg$ satisfies $(+,+)$ orbifold boundary condition\footnote{The boundary condition notation $(\pm, \pm)$ has the meaning that $+\, (-)$ stands for Neumann (Dirichlet) BC of the field wave functions at a brane. The first (second) BC in the listing is at the UV (IR) brane.} because its zero mode is identified as SM gluon. Its mass is then given by
\beq
M_{KK} \, \simeq \, 2.45 \, M_{IR}
\eeq
and will be used to represent the mass scale of KK states. 5D SM fermion will also satisfy $(+,+)$ BC, and its bulk mass parameter $c$ determines its bulk wavefunction, hence its 4D couplings. KK-gluon coupling to SM fermion is shown in \Fig{coup-f0f0g1} \cite{Gherghetta:2000qt}. This vanishes at $c=0.5$ due to orthogonality of wavefuntion solutions. We utilize this property and set $c_{light} =0.5$ to decouple KK-gluons from light fermions. KK-gluons still couple to gluons via triple and quartic vector self-interactions with QCD coupling strength~\cite{Dobrescu:2007yp}.

We comment on our notation regarding the bulk mass parameter $c$. Only one 4D chirality of 5D Dirac fermion will be discussed under orbifold projection. Whether it is LH or RH Weyl, $c=0.5$ is a conformal point where the bulk wave function is flat.

\subsection{$c_{light}=0.5$ and universality}
\label{sec:theory}

We summarize EWPT and the flavor physics of the RS models in general. Then we discuss and review why our assumed parameter space
\beq
c_{light} \= c_b  \= 0.5
\label{eq:c-universal} \eeq
is attractive. In the parameter space of \Eq{eq:c-universal}, $c_b$ is for RH bottom quark, and with \Eq{eq:c-universal} full universality in the RH down sector is achieved.

One can categorize the sources of $\Delta S$ and $\Delta T$: gauge-Higgs, top/KK-fermion, and contributions involving light fermions~\cite{Agashe:2003zs,Carena:2006bn}. A custodial symmetry \cite{Agashe:2003zs,Agashe:2006at} can tame the gauge-Higgs contributions well, and dominant contributors to negative $\Delta T$ are from top/KK-fermion sector which couple strongly via large Yukawa couplings. On the other hand, a global shift of gauge couplings to fermions gives rise to $\Delta S$ ($Zb_L \bar{b}_L$ coupling is well protected by a custodial symmetry). This $\Delta S$ is positive for the commonly assumed case of $c_{light} \sim 0.6$,  conflicting with negative $\Delta T$, which together creates the EWPT tension. Thus decoupling of SM KK gauge bosons from SM fermions can render $\Delta S$ small and make light KK states more viable. $c_{light} \ll 0.5$ will couple too strongly, which may induce too large deviations, while $c_{light} \gg 0.5$ will require too large (non-perturbative) Yukawa couplings. Therefore, $c_{light} \sim 0.5$ is preferentially obtained in many EWPT studies of RS models \cite{Carena:2006bn,Carena:2007ua,Delaunay:2010dw,Cacciapaglia:2004rb,Foadi:2004ps}, if one allows such parameter space which is often not considered in anarchic Yukawa approach.

Universal structure of bulk masses will tame flavor contributions in the RS model. Flavor changing neutral current originating from KK-gluons is induced by generation mixing as well as mixing of zero and KK modes of fermions. The former source can be seen by writing gauge couplings in the mass basis (e.g. RH down sector)
\beq
\tilde{g}_{ij} \= \sum_k \, (D_L)_{ik} \, g_k \, (D_L^*)_{jk} \, \simeq \, (D_L)_{i 3} \, g_3 \, (D_L^*)_{j3}
\label{eq:fcnc-ckm}\eeq
where $D_L$ is some unitary rotation matrix. In the second equation, the equality holds for $c_{light}=0.5 \,(g_1,g_2=0)$, and approximate equality holds even in the usual RS models because $g_3 \gg g_1,g_2$ \cite{Agashe:2004cp}. Thus FCNCs are generically suppressed by small mixing angles whether $c_{light}=0.5$ or not.
In other words, $c_{light}=0.5$ alone without $c_b=0.5$ does not improve the flavor situation significantly from usual RS models, e.g. $\epsilon_K$ still pushes $M_{KK} \gtrsim 20\TeV$ \cite{Csaki:2008zd,Blanke:2009}.

Given the difficulty with flavor physics from third generation mixture, we essentially assume universal bulk masses of the full RH down sector, including $c_b$, as in \Eq{eq:c-universal}. This will get rid of RH down-type FCNCs at leading order since the gauge coupling commutes with the mass matrix in flavor space. Then the bound from $\epsilon_K$ is greatly reduced due to the absence of chiral enhanced LR-type four-fermion FCNC \cite{Csaki:2008zd,Santiago:2008vq}. This kind of universality may be achieved by a bulk flavor symmetry \cite{Cacciapaglia:2007fw,history:flavor-l,Santiago:2008vq} motivated by the AdS/CFT correspondence \cite{ArkaniHamed:2000ds}. We remark that the choice of \Eq{eq:c-universal} actually has been obtained as an attractive solution allowing $M_{KK} \sim 1.5-2\TeV$ in the shining model of ref.\cite{Delaunay:2010dw} with respect to both EWPT and flavor physics, although in their model there is some additional alignment and it is not so important to peg $c_b$ very close to $0.5$. In a somewhat different approach, it has been recognized previously that because $c_b$ plays a less important role than $c_{light}$ in the global fit of EWPT, $c_b \sim 0.5$ is consistently allowed with low KK scale \cite{Carena:2007ua}. It is less required from the data point of view, and perhaps less motivating from the theory point of view, to have $c_{Q,t}$ universal with~\Eq{eq:c-universal}, and thus we generally allow it to vary. However, some of the representative cases that we are studying exploit additional universality in the LH bulk masses (case A and B in the table \ref{sets}) or RH up-type bulk masses (case C and D).

The implications of the mixing of KK and zero mode fermions as a source of FCNC is typically smaller \cite{Blanke:2009}, but one may suspect that pushing light generations closer to the IR brane as in our case may enlarge dangerous mixing effects from Higgs. However, Yukawa couplings become smaller correspondingly, and therefore mixing is reduced. For example, mixing between KK mode $\psi_L^n$ and zero mode $\psi_R^0$ fermions can be written as (by matching 5D Yukawa interaction with 4D effective actions of 4D decompositions $\psi^n(x)$)
\beq
{\cal L}_{yukawa} \, \ni \, \int d^4x \, y H(x) \, \left( f_L^n \overline{\psi}_L^n(x) \right) \, \Big( f_R^0 \psi_R^0(x) \Big) \= \int d^4x \, m_f \frac{f_L^n}{f_L^0} \, \overline{\psi}_L^n(x) \psi_R^0(x)
\eeq
where zero mode fermion mass $m_f = y v \, f_L^0 f_R^0, \, v=\langle H \rangle $ is used. $f^n$ is the $n$-th KK mode fermion bulk profile evaluated at the IR brane. The KK mode profile is always peaked around the IR brane; hence, $f^n$ is almost constant with respect to bulk mass while zero mode $f^0$ is exponentially sensitive as
\beq
f^0 (c) \= \frac{e^{(1/2 - c) \pi k r_c}}{N}, \qquad \frac{1}{N^2} \= \frac{ (1/2 - c) }{e^{(1-2c)\pi k r_c}-1}.
\label{eq:bulk-f} \eeq
Therefore, the $c_{light}=0.5$ case has rather smaller mixing by a factor of $\sim f^0(c=0.6)/f^0(c=0.5) \sim 0.1$. Once the former source of KK-gluon FCNC is well tamed by flavor structure, mixing of KK and zero mode becomes a leading source of down-type FCNC \cite{Santiago:2008vq}. Now the choice of $c_{light}=0.5$ suppresses such well-measured down-type FCNC by additional factors beyond the usual RS case. We comment that smaller 5D Yukawa may increase KK-gluon FCNC in anarchic Yukawa approach. If 5D Yukawas are anarchic, $f^0_{Q,u,d}$'s determine masses, couplings as well as mixing angles, thus giving relations between them. These are given by
\beq
m_{ij} \, \simeq \, v \, y^{5D} f^0_{Li} f^0_{Rj} , \qquad g_i \, \simeq \, g^{5D}\, (f^0_i)^2 \,+ \cdots, \qquad (D_L)_{ij} \, \sim \, \frac{f^0_{Li}}{f^0_{Lj}} \quad \textrm{if } j>i
\eeq
where $\cdots$ represents flavor independent parts of gauge couplings. $y^{5D}$ and $g^{5D}$ are dimensionless couplings of the 5D bulk action in proper units of $1/k$ and $1/\sqrt{k}$, respectively. Then \Eq{eq:fcnc-ckm} becomes
\beq
\tilde{g}_{ij} \, \sim \, f^0_{Li} f^0_{Lj} \, g^{5D} \qquad (i,j<3)
\eeq
which may increase with smaller 5D Yukawa as $f$ can be larger. However, with hierarchic Yukawa, the above relations do not hold and the size of 5D Yukawa coupling is, in general, independent of FCNC gauge couplings.

RS contributions to various dipole operators are relatively suppressed for $c_{light}=0.5$. Dipole operators of the form $\bar{f}_L \sigma^{\mu\nu} f^\prime_R F_{\mu\nu}$ can induce flavor \emph{diagonal} CP violation such as the neutron electric dipole moment (EDM) as well as flavor changing processes such as $b \to s \gamma$ and $\epsilon^\prime / \epsilon_K$ of the $K^0 \to 2\pi$ process. The neutron EDM is typically estimated to be an order of magnitude larger \cite{Agashe:2004cp}, and the other two give one of the strongest bounds on $M_{KK}$ \cite{Agashe:2008uz,Gedalia:2009ws}. Since dipole operators are chirality flipping, dominant contributors are one-loop diagrams in which Higgs and KK fermions are running. These involve at least two (Higgs) mixing insertions of KK-zero mode fermions which give a relative suppression factor of about $\sim (f^0(c=0.6)/f^0(c=0.5))^2 \sim 0.01$ as similarly discussed above.

Smaller Yukawa couplings and consequent smaller KK-zero fermion mixing suppress Higgs-mediated FCNC. In addition, universal $c$ can be capable of additional suppression. If Yukawa couplings between wrong-chirality KK fermions vanish (e.g., between LH SM singlet and RH SM doublet), Higgs FCNC is safely chiral suppressed by $(m_{light}/M_{KK})^2$ because the IR-Higgs boson couples only to 4D chiral modes satisfying the Neumann BC on the IR brane \cite{Blanke:2009,Casagrande:2008hr}. However, Higgs FCNC is generically comparable to KK-gluon induced FCNC with the wrong Yukawa couplings, since wrong Yukawa are generally allowed \cite{Agashe:2009di}. When such wrong Yukawas are identical to corresponding SM Yukawas (which is the case for the bulk Higgs by 5D covariance), universal $c$ eventually has minimal-flavor-violation at low-energy as hierarchical (SM) Yukawa couplings are assumed to be the only flavor spurion (up to non-universal $c_Q$ and $c_t$). Then the leading flavor spurion contribution to Higgs FCNC is aligned with the mass matrices as
\beq
F_Q \, Y_{u,d} Y_{u,d}^\dagger Y_{u,d} \, F_{u,d} \, \sim \, (y_{t,b}^2) \, F_Q \, Y_{u,d}^\dagger \, F_{u,d} \, \propto \,  M_{u,d}
\eeq
because hierarchical $Y_{u,d}^3 \sim (y^2_{t,b}) \, Y_{u,d}$ \cite{Delaunay:2010dw,Agashe:2009di}.  Thus, we are safer with Higgs FCNC as well. We also comment that the degeneracy in $c$ can get rid of radion-mediated FCNC \cite{Azatov:2008vm}.

Another concern of pushing light fermions closer to the IR brane is that higher-dimensional operators composed of light fermions are not suppressed enough purely by fermion localization. The constraint on $\epsilon_K$ requires that relevant dimension-six four-fermion operators need to be suppressed by $\Lambda \gtrsim 10^5\TeV$ while $10^4 \TeV$ is required if CP-phase is absent \cite{Csaki:2008zd}. This effective suppression scale $\Lambda$ is defined by the 4D effective action matched with 5D four-fermion interaction \cite{Gherghetta:2000qt,Cacciapaglia:2004rb}
\beq
{\cal L}_{4D} \, \ni \, \int d^4x \, \frac{a}{\Lambda^2} \, \bar{\psi}^0_i \psi_j^0 \bar{\psi}_k^0 \psi_l^0 \= \int d^4x \int dy \, \sqrt{-g} \, \frac{a}{M_{pl}^3} \, \bar{\Psi}_i \Psi_j \bar{\Psi}_k \Psi_l
\label{four-fermi}
\eeq
where $\psi_i^0$ are zero mode 4D decomposition of 5D Dirac $\Psi_i$. $a$ is a model dependent coefficient. $\Lambda$ is then obtained by integrating products of fermion wave functions over the extra dimension
\beq
\frac{a}{\Lambda^2} \= \frac{a}{M_{pl}^2} \frac{2}{N^4} \frac{ e^{(4-4c) \pi k r_c} -1}{4-4c}
\label{suppscale}
\eeq
with universal $c=c_{light}$. Normalization $N$ is given in \Eq{eq:bulk-f}. Since $c_{light}=0.5$ gives only $\Lambda \simeq 10^2 \TeV$, some sort of cancellation or suppression is necessary which can be achieved by bulk flavor symmetry, for instance. For reference, $\Lambda \simeq 10^5 \TeV$ is obtained for $c_{light} \simeq 0.63-0.67$.\footnote{Values of $\Lambda$ at $c=0.5 (0.65)$ have weak(strong) dependence on the choice of $M_{pl}$ (with fixed $M_{IR}$), as can be seen by computing the ratios $\Lambda(M_{pl}=10^{19} \GeV) / \Lambda(M_{pl}=10^{16} \GeV) \simeq 1.2 (7.9)$ for $c=0.5 (0.65)$. The range for $c_{light}$ quoted in the text is obtained for $k=M_{pl}=10^{19},10^{16}$ GeV substituted into \Eq{eq:ir-k}.}

We express no idealogy as to a theoretical inevitability for $c_{light}=0.5$ besides the data preferring it. We do note that this is a point of enhanced conformal symmetry and therefore could have a strong underlying theoretical motivation. One possible connection may be that when fermions arise from adjoint representations in some D-brane models they are ``born" with $c=0.5$~\cite{Gherghetta:2006yq}.  Strong corrections can push the third generation away from this value, but the light fermions remain there.  Further thoughts on a deeper theory motivation are beyond the scope of this paper.

\subsection{Input parameter ranges}

We fix $c_{light}=0.5$ for the first two generations as assumed, and $c_b=0.5$ for the RH $b$ quark to keep universality of RH down-type bulk masses as discussed.

\begin{figure}
\centering
\includegraphics[angle=0,width=0.42\textwidth]{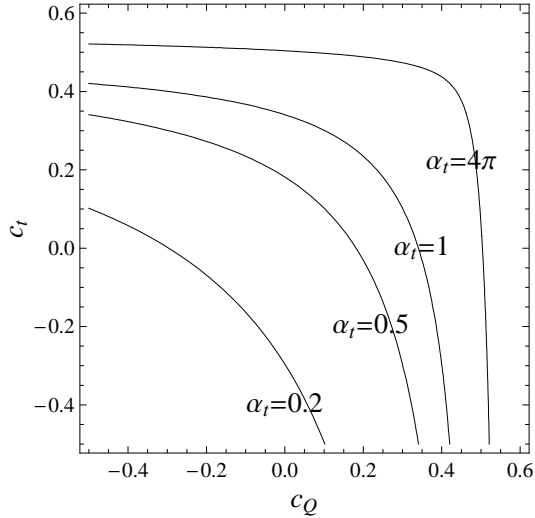}
\caption{Contours of the required 5D top Yukawa $\alpha_t = (y^{5D}_t k)^2/4 \pi$ to obtain top mass $\simeq v$.}
\label{coup-yukawa} \end{figure}

We now have three free parameters. Two of them are remaining bulk masses of third generation. We study the  following range:
\beq
-0.3 \lesssim c_t \le 0.5, \quad -0.3 \lesssim c_Q \le 0.5
\label{eq:c-range}\eeq
for RH top, LH top/bottom doublet, respectively. Both $c_t$ and $c_Q$ are restricted to be less than $0.5$ since otherwise the top Yukawa would be non-perturbative (see \Fig{coup-yukawa}).
Most of the regions of $c_t, c_Q \lesssim 0.5 - 0.4$ allow a perturbative top Yukawa. If $c_Q \lesssim -0.5$, there will be a very light custodian which is excluded (see section \ref{sec:other-sig}), so conservatively we consider the range of $c_Q$ in \Eq{eq:c-range}. Regarding EWPT, a large negative $T$ parameter is typically induced unless $c_t \gtrsim 0.3-0.4$, e.g. $\Delta T = -0.1 \sim -0.15$ for $c_t \lesssim 0.2$  \cite{Carena:2006bn, Carena:2007ua}. For general purposes, we study a rather wide range of $-0.3 \lesssim c_t$ as in \Eq{eq:c-range}, but will focus on cases that are preferred by EWPT (e.g. $c_t \gtrsim 0.3-0.4$ in case C and D in the Table \ref{sets}) when it is relevant.
We refer the reader to section \ref{sec:other-sig} for more discussions on $c_Q$ and $c_t$ with respect to other collider searches.

The last free parameter is the mass of KK gluon $M_{KK}$. We study following range of $M_{KK}$
\beq
1 \TeV \, \lesssim M_{KK} \lesssim 3 \TeV.
\eeq
Previous studies have focused on narrow correlated parameter sets consistent with EWPT that are within the range of such light KK states. However, we study collider phenomenology without being too much restricted a priori by these concerns, but comment on those constraints when relevant.

We choose to present results using specific choices of parameters. Four representative cases that we use are listed in Table \ref{sets}. $M_{KK}$ will be varied within the range above for each of the four cases. We comment that the coupling sets C and D resemble the EWPT-favored parameter space found in ref.\cite{Carena:2006bn,Carena:2007ua}, and even favored by flavor physics in the shining model of ref.\cite{Delaunay:2010dw}. Coupling sets A and B have additional universality structure in LH bulk masses, and case C and D have universal RH up-type bulk masses. These additional structures are not strongly necessary, but may be useful reference points.
\begin{table}[t] \centering \begin{tabular}{|c|c|c|}
\hline
Set A & $g_t = 4$, $g_Q=0$, $g_b=0$ & $c_t \simeq 0.016$ , $c_Q = c_b = 0.5$\\
\hline
Set B & $g_t = 2$, $g_Q=0$, $g_b=0$ & $c_t \simeq 0.305$ , $c_Q = c_b = 0.5$\\
\hline
Set C & $g_t = 0$, $g_Q=3.5$, $g_b=0$ & $c_t = 0.5 $ , $c_Q \simeq 0.1$, $c_b = 0.5$\\
\hline
Set D & $g_t = 0$, $g_Q=2$, $g_b=0$ & $c_t =0.5$ , $c_Q \simeq 0.305$ , $c_b = 0.5$\\
\hline
\end{tabular} \caption{Set of parameters that we use to represent the results. Couplings are in units of $g_{QCD}$. Third column shows values of third family bulk masses that will give corresponding coupling strengths. The first two families are always assumed to have $c_{light}=0.5$.}
\label{sets} \end{table}

\section{Monte Carlo simulation}

\subsection{Signal event generation}
\label{sec:mc-sig}

We have used {\tt MadGraph/MadEvent} v.4.4.42 \cite{Alwall:2007st} for Monte Carlo event generation. CTEQ6M PDF set \cite{Pumplin:2002vw} is used with scale choices of $\mu_R = \mu_F = \sqrt{\hat{s}} / 2$, where $\hat{s}$ is the partonic center of mass energy squared. The KK gluon coupling strength is assumed to run according to the QCD beta function. NLO corrections are not included. Since a KK gluon is a broad resonance, the narrow width approximation (on-shell production and subsequent decay) is not a good approximation. For light $M_{KK}$ {\tt MadEvent} produces $5 - 20\%$ difference between cross sections from on-shell-and-decay and full matrix element computations, but a rather large difference of up to $\sim {\cal O}(100 \%)$ for heavy KK states $M_{KK} \gtrsim 2 - 2.5 \TeV$. We have generated the full matrix elements of $\tT \tT$ and $\tT b\bar{b}$ production in {\tt MadGraph/MadEvent} to take into account broad resonance effects, and we decay top quarks using {\tt BRIDGE} \cite{Meade:2007js}. All results are obtained for $\sqrt{s} = 14 \TeV$ LHC. We assume a luminosity ${\cal L} = 100\xfb^{-1}$ when a value is necessary.

\begin{figure}
\centering
\includegraphics[angle=0,width=0.48\textwidth]{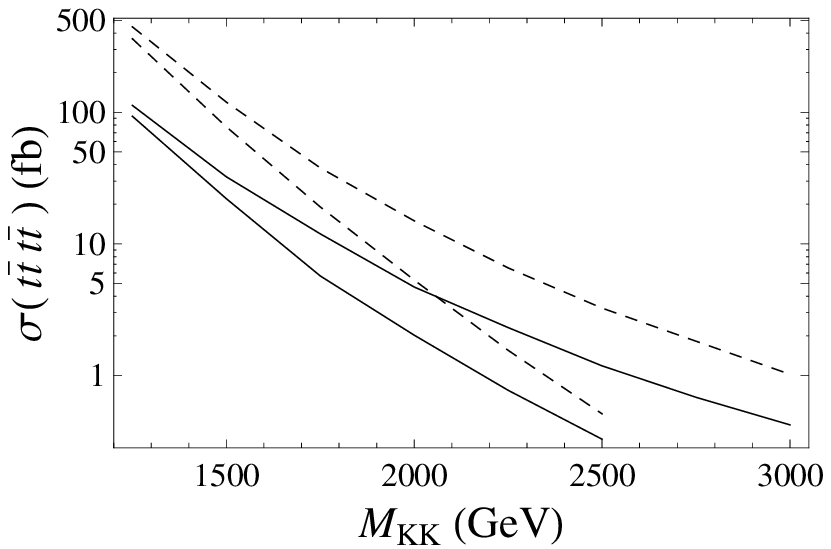}
\includegraphics[angle=0,width=0.48\textwidth]{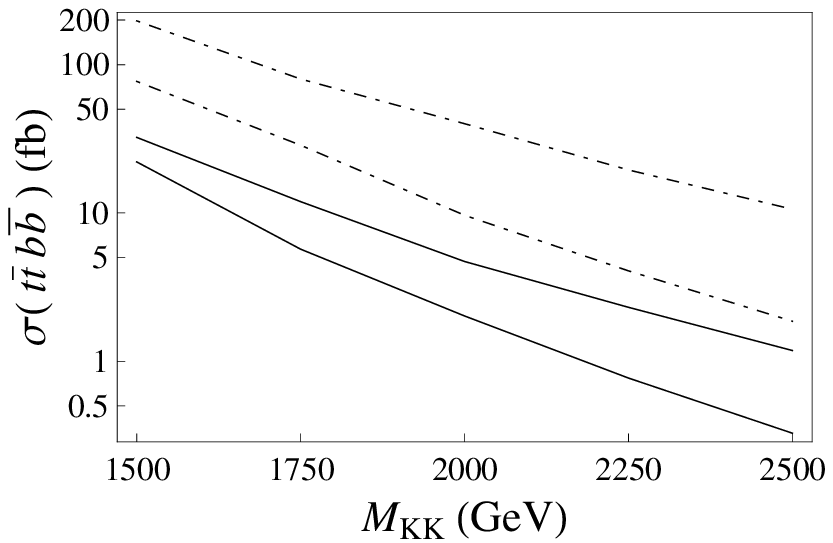}
\caption{(Left): $\sigma(pp \to \tT\tT)$ for coupling set A,B(dashed lines) and C,D(solid lines). (Right): $\sigma(\tT b\bar{b})$ for set C(upper dot-dashed) and set D(lower dot-dashed). Corresponding $\sigma(\tT \tT)$ are also shown as solid lines for comparison. Coupling sets can be found in Table \ref{sets}. $\sqrt{s} = 14 \TeV$ at LHC. }
\label{cross-section} \end{figure}

Cross sections of $pp \to \tT\tT,\, \tT b\bar{b}$ via $\KKg$ are plotted in \Fig{cross-section}. Representative cases of $\KKg$ interacting only with RH top (dashed lines), and $\KKg$ interacting only with LH top (solid lines) are shown. One qualitative feature that this plot shows is that cross sections in the light $M_{KK}$ region are governed by vector self-interactions (between gluons and KK gluons). The dependence on fermion couplings is mostly in the branching ratio $Br(\KKg \to \tT)$. Two dashed lines (solid lines) have the same $Br(\KKg \to \tT) = 100\%\, (50\%)$ \footnote{We ignore KK-gluon decays into KK-fermions because the parameter space with heavy KK-fermion is our interest.} and thus they approach a common value in the light $M_{KK}$ limit. However, fermion coupling dependence becomes important in the heavy $M_{KK}$ region because vector self-interactions contribute only to pair production of $\KKg$ that drops more quickly than the associated production.

The other notable feature in \Fig{cross-section} is the effect of the bottom coupling in $\tT \tT$ production. Bottom coupling changes the branching ratio $Br(\KKg \to \tT)$ which shows up in two ways: reduction of the total rate, and weaker dependence on $M_{KK}$. Both effects can be observed by comparing case B (lower dashed) and case D (lower solid) lines in \Fig{cross-section}. They have the same size of top coupling; one is LH (solid) and the other is RH (dashed). However, the LH case has a lower rate of four top due to smaller branching ratio into $\tT$. More interestingly, two lines behave differently with $M_{KK}$: the RH case (dashed) drops more quickly with $M_{KK}$. The smaller branching $Br(\KKg \to \tT)$ of the LH case will suppress pair production ($\KKg \KKg \to \tT \tT$) more relative to the associated production ($\tT \KKg \to \tT \tT$). As pair production phase space quickly becomes smaller with $M_{KK}$, the LH case will have weaker dependence on $M_{KK}$.

$\tT b\bar{b}$ events are also produced if the LH top coupling is turned on (i.e. $g_Q \ne 0$). This can contribute to single lepton and opposite-sign diletpon final states. The rate is usually higher than four top production by a factor of $4-8$ as $b\bar{b}$ phase space is larger than $\tT$ (see \Fig{cross-section}). However, leptons in $\tT b\bar{b}$ generally must come from $\tT$ pair and this topology resembles that of the main SM background $\tT$. Indeed, it turns out that the majority of $\tT b\bar{b}$ events are cut out by event selections efficient for SM $\tT$ background reduction. For the single lepton final state, the $\tT b\bar{b}$ contribution is larger than the $\tT \tT$ contributions only by a factor of $1-2$, so do not lose much discovery capability.

One concern about $\tT b\bar{b}$ is the reliability of the Monte Carlo cross section calculation. The dominant contribution to $\tT b\bar{b}$ comes from $b\bar{b} \KKg$ associated production in which a large theoretical uncertainty of $b\bar{b}$ cross section may be present. Theoretical uncertainties originate from possible small scale choices, log enhanced IR contribution from collinear $b$ production, etc. Thus this process should be studied more carefully once we can normalize the Monte Carlo simulation properly with real data. Here, we simply take minimal cuts on the $b\bar{b}$ pair to partially avoid such complications. Those are $\Delta R(b,\bar{b}) \ge 0.1, \, p_T(b) \ge 10 \GeV, \, m_{b\bar{b}} \ge 10\GeV$.

\subsection{Background study}

Many dedicated studies of multi-lepton final states have been carried out in the literature. Our strategy is to use some of those available results and compare our signals with them to estimate discovery potential of the four top channel. Most background studies are aimed at heavy resonance searches as in our case. We shall see that some common features of heavy new physics enable our four top signals to beat SM backgrounds.
By working with many different background studies suited for different models, we will be able to capture important qualitative features of our four top signals. We comment on some useful different features of the four top signals when it is appropriate.

Multi-lepton background results used in this paper are from several supersymmetry searches probing different parts of the parameter space \cite{Baer:1995va,Aad:2009wy,Acharya:2009gb}, searches of exotic fermions coupling to the third generations \cite{AguilarSaavedra:2009es,Contino:2008hi}, search of the light Higgs in the $WH \to WWW$ channel \cite{Cavasinni:2000,Baer:1998cm}, and search of compositeness of top quark \cite{Lillie:2007hd}. Among them, ref. \cite{Baer:1995va,Aad:2009wy} are main sources that we use to analyze our results.

We do not carry out a fully detailed collider study including detector effects, optimizing cuts, and studying fake or mis-measurement ratios. That will be done at the appropriate time by the experimental groups after understanding LHC detectors well with real data. Further optimizations of the event selection are not done here because signal cross section is already small ($\sim 10\, {\rm fb}$), around the upper limit of discovery reach.

\section{Like-sign dilepton topology}

Like-sign dilepton (LSDL) final states are defined as two and only two leptons (i.e., electrons or muons, but not tau leptons) with same charge accompanied by sizable missing energy. This is quite rare in the SM, which makes it one of the most promising signals of four top quark production when the goal is restricted to the first stage of just determining if there is beyond the SM processes at work in the data.  We shall describe the background contributions and estimate the prospects of detecting the beyond the SM signal.

\subsection{Comparison with background studies from supersymmetry searches}

Following ref. \cite{Baer:1995va}, we employ the cuts:
\begin{itemize}
\item LSDL event selection set $\#$ 1:
 \begin{enumerate}
 \item \underline{only LSDL} with $p_T \geq 20 \GeV, \, |\eta| \leq 2.5, \, \Delta R_{lj} \geq 0.3$
 \item \underline{at least two jets} with $p_T \geq E_T^c, \, |\eta| \leq 3.0$, no b tagging
 \item $E_T^{miss} \geq E_T^c$
 \end{enumerate}
\end{itemize}
These cuts are optimized for supersymmetry searches, but we find them reasonably well optimized for the four top signatures we have in mind here.
$E_T^c$ is a useful variable that is to be varied to see the discovery reach because new physics contributions are likely to be higher $p_T$ than SM backgrounds. Leptons are from long decay chains, and thus mild $p_T$ cuts are used. Note that $\Delta R_{lj} \geq 0.3$ is not the standard isolation cut ($\Delta R_{lj} \geq 0.4$) that is used commonly now by ATLAS, CMS groups (e.g. see ref.\cite{Aad:2009wy}). This will slightly overestimate the potential of LSDL observable as can be deduced from Table \ref{tbl-toplep}. Also, $|\eta_j| \leq 3.0$ is not likely to be the standard choice ($2.5$), but this modification is very insignificant since high $p_T$ objects are generally very central.

\begin{figure}
\centering
\includegraphics[angle=0,width=0.48\textwidth]{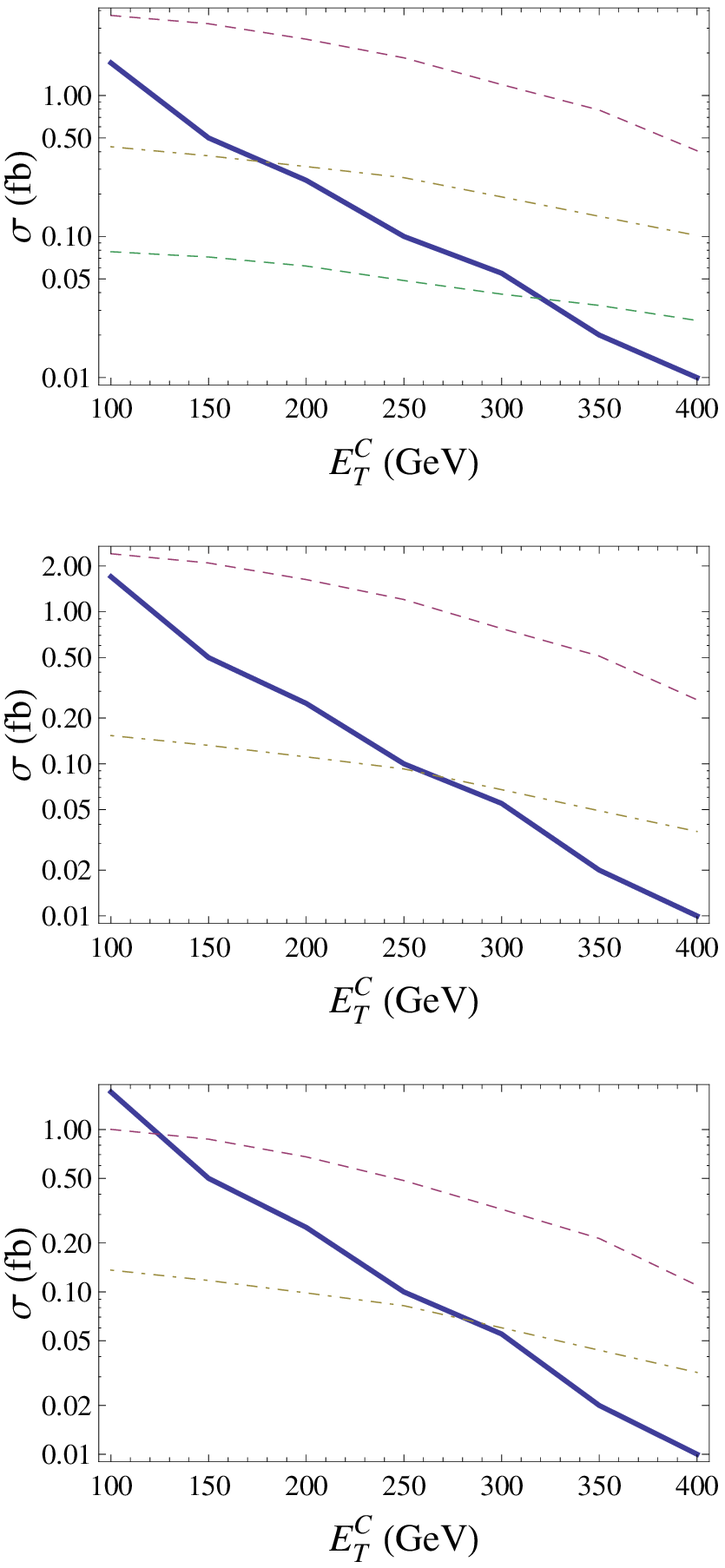}
\includegraphics[angle=0,width=0.48\textwidth]{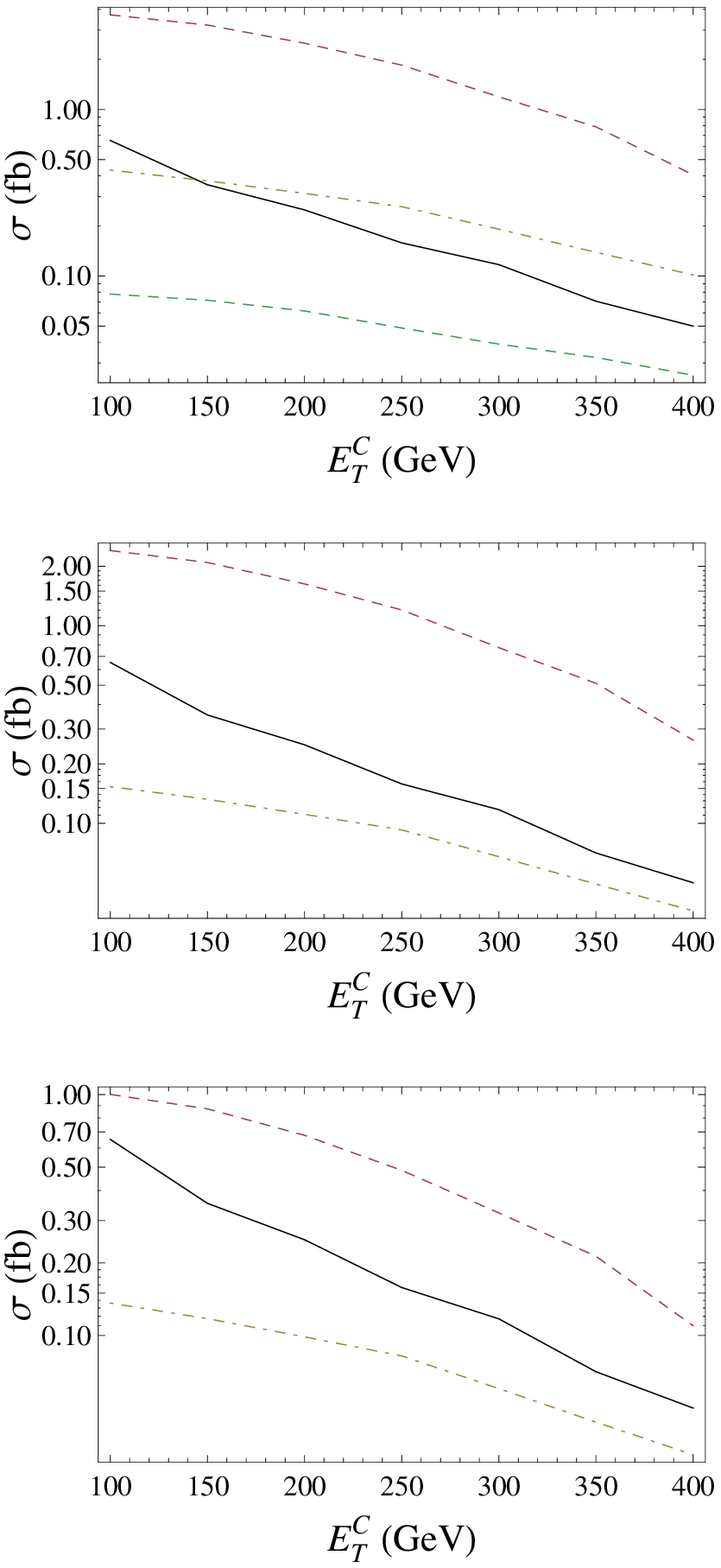}
\caption{LSDL results for case A,B,C from top to bottom. (Left): LSDL cross sections after LSDL $\#1$ cuts. Background (thick solid) from ref.\cite{Baer:1995va} and signal for $M_{KK}=2000\GeV$ (dot-dashed), $1500,\,2500\GeV$ (dashed). (Right): Same plot as the left panel, but the corresponding $5\sigma$ reach with $100\xfb^{-1}$ of data is shown as a solid line.}
\label{res-lsdl} \end{figure}

Cross section results of signal and background (from ref. \cite{Baer:1995va}) are shown in \Fig{res-lsdl}. The $5\sigma$ discovery reach of the cross section is also shown in the right panel by using
\beq
\frac{S}{\sqrt{B}} \= \frac{ {\cal L} \cdot \sigma_{signal} }{ \sqrt{ {\cal L} \cdot \sigma_{bkgd} } } \, \geq \, 5
\eeq
where the luminosity ${\cal L} = 100\xfb^{-1}$ is assumed.

For the strong coupling case A, $M_{KK} \simeq 2200 \GeV$ can be probed with optimized choice of $E_T^c \simeq 250 - 350 \GeV$ while $M_{KK} \lesssim 1700 \GeV$ is accessible for the weaker coupling cases B and C. Since the background drops more quickly than signal with $E_T^c$, discovery reach with lower $E_T^c$ value is smaller. For higher values of $E_T^c$, the issue is then small number of signal events rather than suppressing background. If we require having at least 10 signal events (equivalent to $\sigma_{signal} = 0.1$ fb after cuts) to claim evidence of new physics, $E_T^c \gtrsim 300 - 350 \GeV$ should not be taken for heavy $M_{KK} \sim 1.7 - 2 \TeV$. However, a simple ratio of signal and background $\sigma_{signal}/\sigma_{bkgd}$ increases even for very high $E_T^c \gtrsim 350 \GeV$ for all cases (see the left panel).

In the LSDL channel, either lepton must come from a $\KKg$'s daughter top. This top is likely to be boosted, and its lepton is less likely to be isolated from jetty activities. We study this lepton-jet collimation issue in detail in section \ref{sec-leptop}.

\subsection{Comparison with other LSDL studies}

We compare our signal with several other background studies suited for different models. Two are from gluino pair production \cite{Aad:2009wy, Acharya:2009gb}, one from four-top composite operator \cite{Lillie:2007hd}, and the other from pair production of heavy exotic quarks decaying to $tW$ \cite{Contino:2008hi}. Kinematic cuts are varied between those references, and they seem to consider somewhat different sets of background processes. However, final backgrounds after cuts are all about $3-7 \fb$. So we simply compare our signal cross sections with this value of background cross section.

We impose the following kinematic cuts that resemble the strongest set of cuts among the references listed above:
\begin{itemize}
\item LSDL event selection set $\#$ 2:
 \begin{enumerate}
 \item \underline{jet}: leading $p_T \geq 100\GeV,\, p_T \geq 80\GeV, \, |\eta|\leq 5$, and $n_j \geq 4$, no b-tagging.
 \item \underline{only LSDL lepton}: leading $p_T \geq 50\GeV, \, p_T \geq 25\GeV, \, |\eta|\leq 2.5, \, \Delta R_{lj} \geq 0.4$
 \item $E_T^{miss} \geq 20\GeV$.
 \end{enumerate}
\end{itemize}
Note that multi-jet requirement may underestimate our signal sample because our sample does not take into account initial/final state radiations. We have tried to vary number of jet requirement and checked that selection efficiency changes by a factor of $\sim 2$ within $n_j \geq 3-5$.

Event selection efficiency for several values of $M_{KK}$ are given in Table~\ref{lsdl2}. Four-top cross sections multiplied by efficiency and branching ratios are also shown. Given the background cross sections of about $3-7\fb$, $100 \fb^{-1}$ of data can achieve $5 \sigma$ significance if the signal cross section after cuts is greater than about $0.9 - 1.3 \fb$. So it is likely that $M_{KK} \lesssim 1600 \GeV$ for coupling set A can be probed with event selection $\#2$ using much milder $p_T$ cuts than $\#1$.

\begin{table}[t] \centering \begin{tabular}{|c|c|c|c|c|c|}
\hline
$M_{KK}$ (GeV) & 1000 & 1500 & 2000 & 2500 & 3000 \\
\hline \hline
efficiency of LSDL $\#$2 cuts & 4.3 $\%$ & 2.8 $\%$ & 2.3 $\%$ & 1.9 $\%$ & 1.3 $\%$ \\
\hline
$\sigma(\tT \tT)_A \cdot$ efficiency (fb) & $\sim 90$ & $\sim 2.5$ & $\sim 0.3$ & $\sim 0.06$ & $\sim 0.01$ \\
\hline
\end{tabular} \caption{LSDL results. Efficiencies (including branching ratios) under cuts $\#2$, and signal cross sections after cuts for case A is also shown.} \label{lsdl2} \end{table}

\section{Single lepton final states}

The single-lepton observable ($1l$ + high $p_T$ jets + $E_T^{miss}$) is studied in this section. Later in the section, we briefly study multi $b$-tagging method, instead of requiring high $p_T$ objects which seems to be more suited for supersymmery searches.

\subsection{Discovery potential}
Following ref. \cite{Baer:1995va} we use following cuts:
\begin{itemize}
\item Single lepton event selection $\#1$:
  \begin{enumerate}
    \item \underline{only one lepton} with $p_T \geq 20 \GeV, \, \Delta R_{lj} \geq 0.3, \, |\eta| \leq 2.5$
    \item \underline{at least two jets} with $ p_T \geq E_T^c, \, |\eta| \leq 3.0$
    \item $E_T^{miss} \geq E_T^c, \quad M_T(l,E_T^{miss}) \geq 100 \GeV$
  \end{enumerate}
\end{itemize}
where the transverse mass $M_T$ is defined using transverse four-vectors of a lepton $p_T^l$ and transverse missing energy $p_T^{miss}$ (treated as a light-like four vector pointing perpendicular to the beam axis). It is defined as
\beq
M_T^2(l,E_T^{miss}) \= 2(E_T^l E_T^{miss} - p_T^l \cdot p_T^{miss}).
\eeq
Transverse mass will be around the $W$ boson mass if the lepton and missing energy are from a single $W$ boson, hence effectively suppressing SM backgrounds with missing energy from $W \to l \nu$. This will also suppress the contribution from KK gluon mediated $b\bar{b} \tT$ since leptons generally must come from $\tT$ as discussed in section \ref{sec:mc-sig}.

We comment that if we modify the old lepton isolation criteria used here to the more standard one (i.e. $\Delta R_{lj} \geq 0.3 \to 0.4$), event selection efficiency is not necessarily reduced. This is simply because many multi-lepton events ($n_l \geq 2$) will now have higher chances to contribute to the single lepton event samples with tighter lepton isolation. Refer to section \ref{sec-leptop} for discussions regarding lepton isolation and lepton-jet collimation issues.  This modification brings only about a ${\cal O}(0.1)\%$ efficiency change. Also, the modification $|\eta_j| \leq 3.0 \to 2.5$ introduces negligible changes.

\begin{figure}
\centering
\includegraphics[angle=0,width=0.48\textwidth]{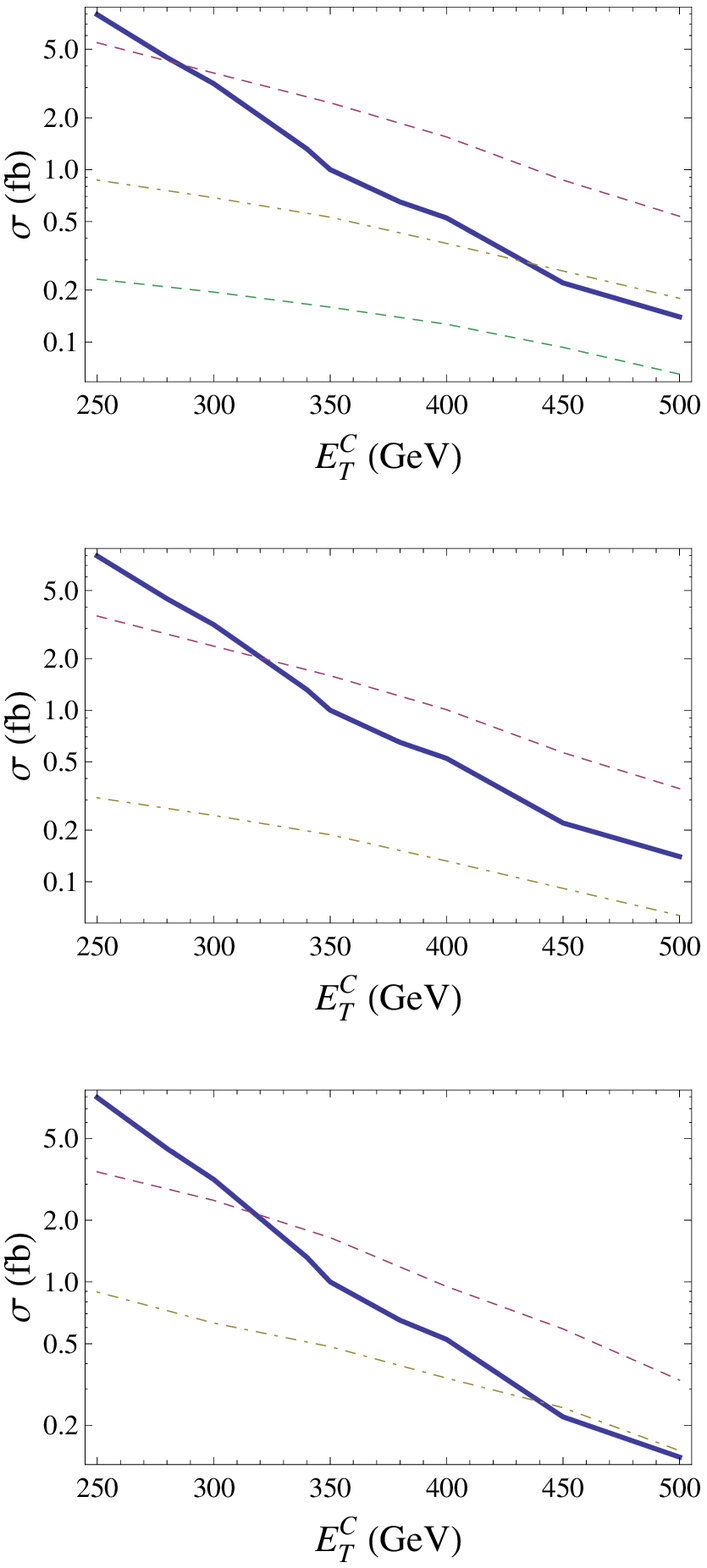}
\includegraphics[angle=0,width=0.48\textwidth]{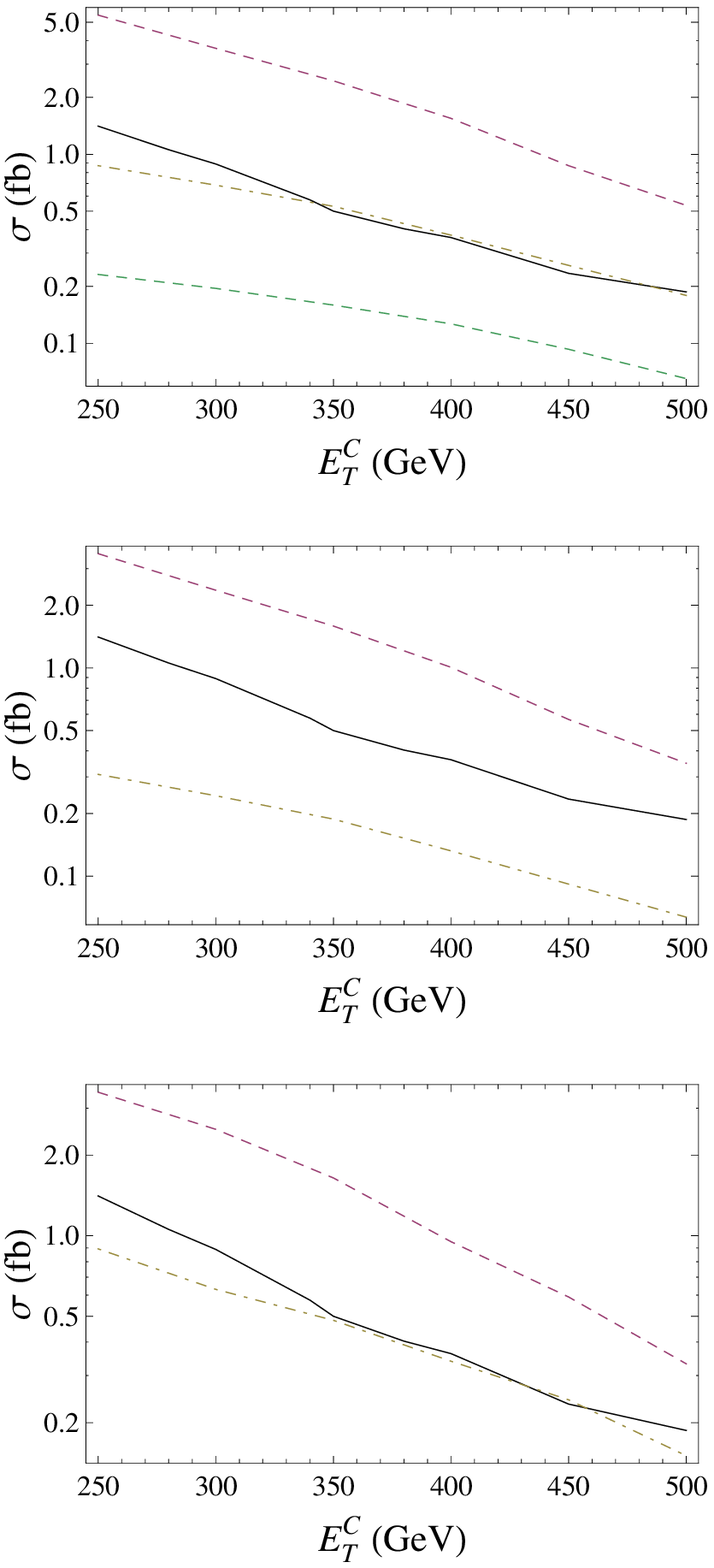}
\caption{Single lepton results for case A,B,C from top. (Left) Single lepton cross sections after cut $\#1$. Background (thick solid) from ref.\cite{Baer:1995va} and $M_{KK}=2000\GeV$ (dot-dashed), $1500,\,2500\GeV$ (dashed). (Right) Same plot as the left panel, but the $5\sigma$ reach with $100\xfb^{-1}$ of data is shown as a solid line. }
\label{res-singlep} \end{figure}

Cross section results are shown in \Fig{res-singlep}. The strong coupling case A can be probed up to $M_{KK} \simeq 2000 \GeV$ with $E_T^c \gtrsim 350 \GeV$ while case B has lower discovery reach of about $M_{KK} \lesssim 1700 \GeV$. Case C can reach $M_{KK} \simeq 1950 \GeV$ with $E_T^c \gtrsim 350 \GeV$, which is better than what can be obtained in case B. On the other hand, in the like-sign dilepton and trilepton searches, case C has a lower discovery reach than case B (see \Fig{res-lsdl} and \ref{res-trilep}). This better discovery potential of case C here is due to the larger contributions from $\tT b\bar{b}$ that only exists for case C and D. By the same reason, we will see that case C has the highest discovery potential in the single lepton channel while case A and B typically do not.

It is interesting to compare the above results of cut $\#1$ with results obtained by very similar cuts used by the ATLAS Supersymmetry group (p.1597 of ref.\cite{Aad:2009wy}). Those new sets of cuts are
\begin{itemize}
\item Single lepton event selection set $\#2$:
\begin{enumerate}
\item \underline{only one} lepton with $\quad p_T \geq 20 \GeV , \quad \Delta R_{lj}\geq 0.4 , \quad |\eta|\leq 2.5$
\item no additional leptons with $\quad p_T \geq 10 \GeV$
\item \underline{at least} 4 jets with $\quad p_T \geq 50 \GeV$, and leading $p_T \geq 100 \GeV$. $|\eta| \leq 2.5$
\item $E_T^{miss} \geq max( 100 \GeV,\, 0.2 H_T)$. For heavy resonances, always $0.2 H_T > 100 \GeV$.
\item $M_T(l,E_T^{miss}) \geq 100 \GeV, \quad H_T \geq 800 \GeV, \quad S_T \geq 0.2$
\end{enumerate}
\end{itemize}
Cross section results are shown in the Table~\ref{res-singlep2}. The $5\sigma$ discovery reach of cuts $\#2$ is $\sigma(\tT\tT) \cdot eff \gtrsim 3.2$ fb \cite{Aad:2009wy} that is interpreted as $M_{KK} \lesssim 1600 \GeV$ for set A from the Table~\ref{res-singlep2}. This is a lower reach than what can be obtained using cuts $\#1$.

Cuts $\#2$ resemble cuts $\#1$ in the sense that high $E_T^{miss}$ and $M_T$ cuts are used to suppress backgrounds. However, the high $p_T$ jet requirement ($p_T \gtrsim 350 \GeV$) in set $\#1$ is replaced by high multiplicity jet topology ($n_{jet} \ge 4$) in set $\#2$, and the former wins in our case. High $p_T$ objects are very useful probes of new physics beyond the SM as usually expected.

\begin{table}[t] \centering \begin{tabular}{|c|c|c|c|c|c|}
\hline
$M_{KK}$ (GeV) & 1000 & 1500 & 2000 & 2500 & 3000 \\
\hline \hline
efficiency of single lepton $\#$2 cuts & 3.1 $\%$ & 3.6 $\%$ & 3.9 $\%$ & 4.5 $\%$ & 5.2 $\%$ \\
\hline
$\sigma(\tT \tT)_A \cdot$ efficiency (fb) & 68 & 4.3 & 0.58 & 0.15 & $\sim 0.05$ \\
\hline
\end{tabular} \caption{Single lepton results. Efficiencies(including branching ratios) of event selection $\#2$, and case A signal cross sections after cuts are shown.} \label{res-singlep2} \end{table}

\subsection{Three $b$-tagging method}

In the discussion above we have found to be true the expected qualitative result that utilizing high $p_T$ jets is very useful in the search of new physics. Four top events in RS models is no except, and we can take advantage of high $p_T$ cuts on jets and missing energy just like in supersymmetry as we saw in the previous subsection. However, RS four top events have qualitative differences from supersymmetry events.

RS four top events via KK-gluons have generically smaller missing energy than supersymmetry events. Cascade decays of supersymmetry particles typically end up with energetic neutrinos from heavy particle decays and/or heavy LSPs. These give rise to large missing energy. RS also predicts heavy particles decaying to neutrinos (e.g. KK excited W boson). However, decay chains of KK particles in our scenario quickly ends up only with SM particles, thus the majority of neutrinos are from $W$ boson decay, with ultimately somewhat softer missing energy spectrum.
Moreover, most leptons in RS four-top events are from $W$ boson decays (not from heavy particle decays such as gauginos). Although the $W$ bosons could be boosted somewhat by being daughter particles of heavier KK state production, the $M_T$ between a lepton and missing energy will nevertheless be more likely to be around the $W$ boson mass than in the supersymmetry case.

Given these observations, we study an alternative single lepton observable. We impose rather milder cuts of $E_T^{miss}$ and $M_T$, and additionally a three $b$-tagging requirement (to suppress backgrounds):
\beq
1l \,+ \, \geq 3 \textrm{ b-tagged jets} \, + \, \textrm{mild $p_T$ jets, } \, E_T^{miss}
\label{singlep-3b} \eeq
The simultaneous existence of a single isolated lepton and three $b$-jets are rare in the SM. We comment that a $b$-tagged jet does not have to be in reality a $b$-jets, but can also be a (boosted) top-jet which can increase discriminating power. Similar observable has indeed been used to search supersymmetry in ref.\cite{atlas:3b} based on the default ATLAS $b$-tagging algorithm (suited for non-boosted $b$-jets), and it has resulted in SM background of ${\cal O}(100)$fb which might still be too large for our four-top signal \cite{atlas:3b}. We use following event selections of the alternative observable
\begin{itemize}
\item Single lepton event selection cuts $\#3$:
\begin{enumerate}
\item Definitions of jet, lepton from cuts 1,2,3 of event selection $\#2$.
\item \underline{At least} 3 b-tagged jets, and $b$-tagging efficiency $\epsilon_b$ will be varied.
\item $E_T^{miss} \geq 100 \GeV$, \underline{No} $M_T$ cuts, $H_T \geq 1000 \GeV$.
\end{enumerate}
\end{itemize}
Optimization of this alternative observable using three $b$-taggings, and consequent comparison with the previous results using high $p_T$ objects shown in \Fig{res-singlep} are interesting, but will be a future project. High cuts on $H_T$, scalar sum of $p_T$ of all objects in the event, is almost harmless for signal \cite{Lillie:2007hd,Mrazek:2009yu}.

Major backgrounds are categorized in Table \ref{bkgd:3b}.
\begin{table}[t] \centering \begin{tabular}{|l l l|}
\hline
 & Backgrounds & Cross sections \\
\hline
\hline
(a) & $W/Z +$ jets, $WZ +$ jets & $\sigma \simeq 8 \times 10^7\xfb$ with $p_T(j) \geq 10 \GeV$ \\
(b) & $\tT +$ jets, $W/Z + b\bar{b}$ & $\sigma \simeq 9 \times 10^5\xfb$ \\
(c) & $b\bar{b}$, $\tT b\bar{b}$ & $\sigma(b\bar{b}) \simeq 2 \times 10^8\xfb$ with $p_T(b) \geq 40 \GeV$\\
\hline
\end{tabular} \caption{Categorization of major backgrounds to single lepton observable in \Eq{singlep-3b}. Background numbers are taken from ref.\cite{Cavasinni:2000} (see also refs.\cite{atlas:3b,AguilarSaavedra:2009es}).} \label{bkgd:3b} \end{table}
Category (a) have $\geq 1$ lepton + no $b$-jets, category (b) have $\geq 1$ lepton + $2$ $b$-jets, and category (c) are none of these. Due to small number of $b$-jets in category (a), highly efficient $b$-tagging of rejection about $\sim 200-400$ can suppress backgrounds (a) below $\sim 0.1$ fb (where 10 events are obtained with $100\xfb^{-1}$) with a reasonable value of event selection efficiency times branching ratio $\sim {\cal O}(1 \%)$. To suppress (b), which already have several true $b$-jets, one may need to require a certain number of \emph{boosted} objects (e.g. top) because $b$-jets from SM top quarks may resemble QCD background jet characteristics more than a boosted object. If a boosted jet-tagging algorithm can obtain a powerful rejection factor of $\sim 200-700$ and if tagging of two boosted objects is required, category (b) can be negligible. For $b\bar{b}$ in (c), by assuming jet-lepton faking rate of $\sim 10^{-4} - 10^{-5}$ and the probability of isolated leptons from leptonic decay of a $b$-jet about $\sim 10^{-5}$ (e.g., see ref.\cite{Aad:2009wy,Cavasinni:2000}), expected additional rejection (from lack of three $b$ taggings) of greater than $\sim 100$ is enough to get rid of $b\bar{b}$ backgrounds. The small cross section $\sim 1\, {\rm pb}$ of SM $\tT b\bar{b}$ may render this background negligible even though its event topology resembles some of our signal. We note that such a highly efficient (boosted) jet-tagging algorithm desired has been discussed in ref.\cite{Rehermann:2010vq} in the context of boosted leptonic top tagging.
Rather than estimating these backgrounds more accurately, which are subject to large uncertainties (fakes, mis-measurements, etc.), we simply study the discovery reach as a function of $b$-tagging efficiency $\epsilon_b$ with our reasonably assumed small backgrounds of $\lesssim 0.1\xfb$.

Results are shown in Table \ref{res-singlep3} for several values of $b$-tagging efficiency $\epsilon_b$. With our given estimate of backgrounds, which may be optimistic compared to what a full experimental study would conclude, discovery reach is quite high around $3\, {\rm TeV}$. Given that this may be one of the best signatures, it would be interesting for experiments to carry out a full simulation to compute precisely the rejection factor from lepton isolation and also the true $b$-tagging efficiency for $M_{KK} \gtrsim 2 \TeV$.

\begin{table}[t] \centering \begin{tabular}{|c|c|c|c|c|}
\hline
Single lepton & $\epsilon_b = 40\%$ & $\epsilon_b = 50\%$ & $\epsilon_b = 60\%$ & $\epsilon_b = 100\%$ \\
\hline
$M_{KK} =$ 1500 GeV & 2.7$\%$ (1.4$\%$) & 4.8 (2.5)  & 7.5 (3.9)  & 20.3 (11.3) \\
2000 GeV & 2.6 (1.3) & 4.7 (2.3) & 7.3 (3.7) & 19.9 (11.5) \\
2500 GeV & 2.5 (1.0) & 4.4 (1.8) & 6.9 (2.9) & 18.9 (8.8) \\
3000 GeV & 2.4 (0.8)  & 4.3 (1.5) & 6.8 (2.4) & 18.8 (7.5) \\
\hline
10 signal A events &  2400 GeV & 2600 & 2800 & $\sim 3200$ \\
10 signal C events &  2500 GeV & 2700 & 2900 & $\sim 3200$ \\
\hline
\end{tabular}
\caption{Efficiencies with single lepton selection $\#3$ utilizing three $b$-tagging. $b$-tagging efficiency $\epsilon_b$ is varied. Results of $\tT\tT$ ($\tT b\bar{b}$) event samples are shown, respectively. Maximum $M_{KK}$ giving rise to 10 signal events at ${\cal L} = 100\, {\rm fb}^{-1}$ is also shown for points A and C.} \label{res-singlep3}
\end{table}

\section{Trilepton search}

The trilepton observable is three charged leptons of any charges plus either $E_T^{miss}$ or high $p_T$ jets. All events have either $(++-)$ or $(--+)$ charge combinations in our case since the total charge of two colliding partons do not exceed $\pm 1$. We include both charge combinations.

\subsection{Discovery potential}

\begin{figure}
\centering
\includegraphics[angle=0,width=0.48\textwidth]{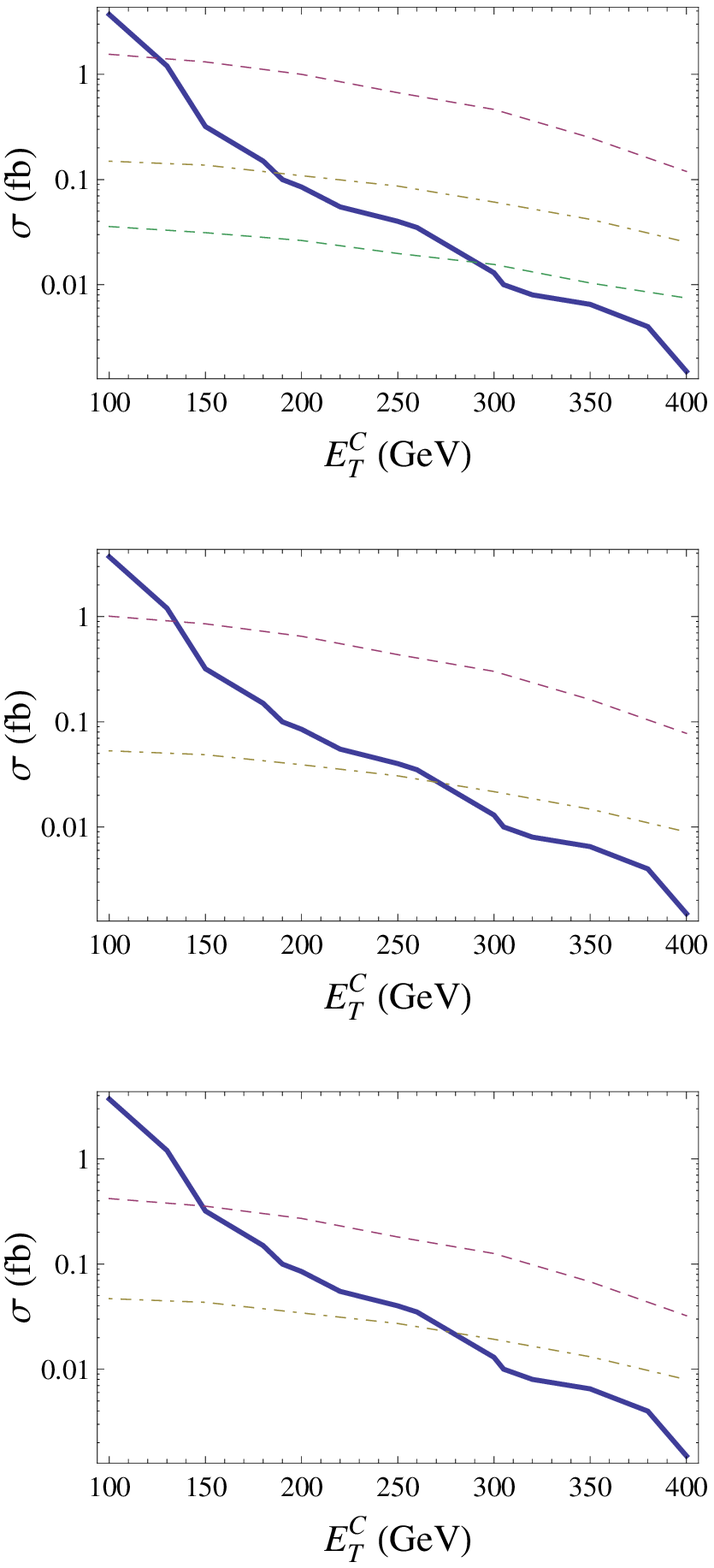}
\includegraphics[angle=0,width=0.48\textwidth]{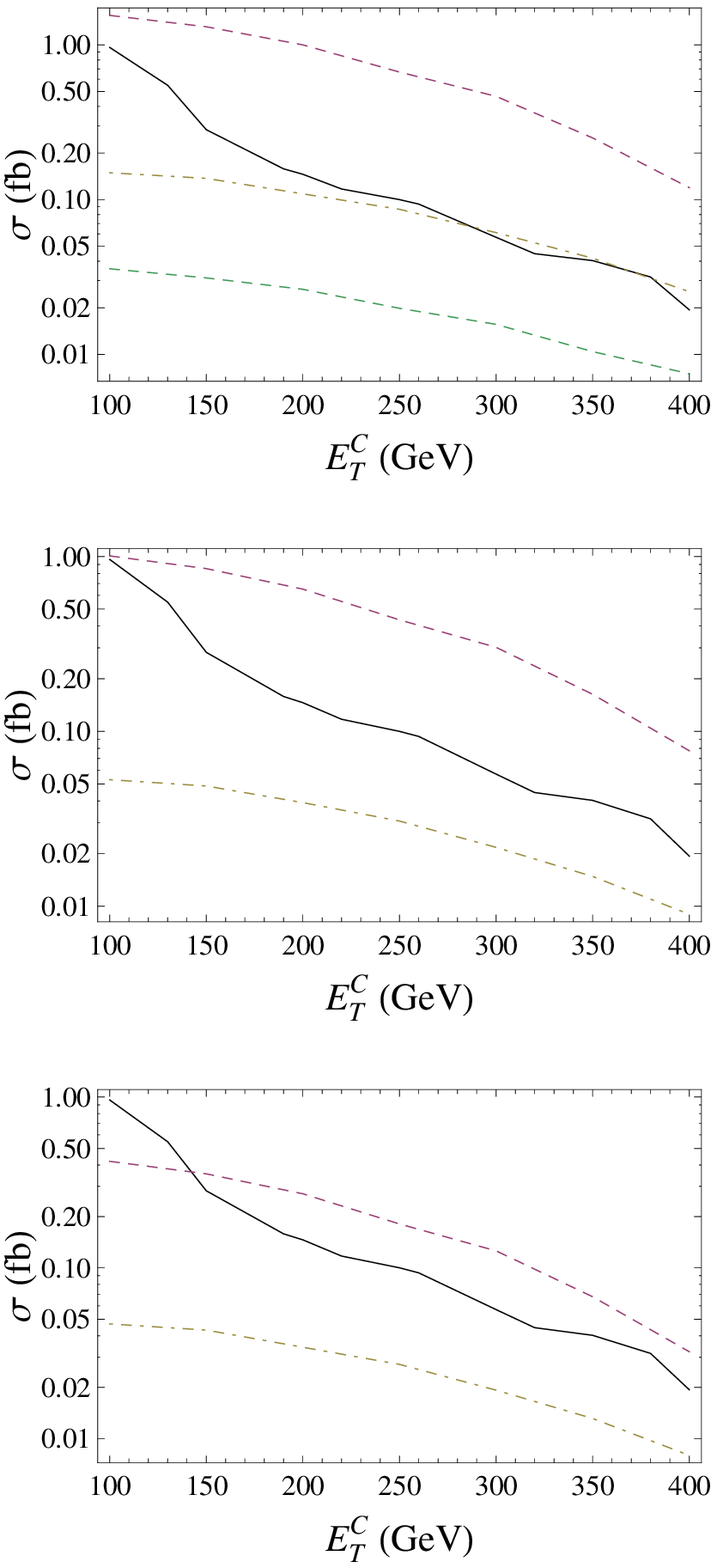}
\caption{Tri-lepton results for case A,B,C from top to bottom. (Left) Tri-lepton cross sections after Tri-lepton cuts $\#1$. Background (thick solid) from ref.\cite{Baer:1995va} and $M_{KK}=2000\GeV$ (dot-dashed), $1500,\,2500\GeV$ (dashed). (Right) Same plot as the left panel, but the $5\sigma$ reach with $100\, {\rm fb}^{-1}$ of data is shown as a solid line. }
\label{res-trilep} \end{figure}

First, we use cuts in ref.\cite{Baer:1995va}, which  were originally employed to reduce backgrounds for supersymmetry searches:
\begin{itemize}
\item Tri-lepton event selection set $\#1$:
\begin{enumerate}
\item \underline{Only 3} leptons with $p_T \geq 20\GeV,\, |\eta| \leq 2.5,\, \Delta R_{lj}\geq 0.3$
\item \underline{At least 2} jets with $p_T \geq E_T^c,\, |\eta| \leq 3.0$, no $b$-tagging
\item $E_T^{miss} \geq E_T^c$
\end{enumerate}
\end{itemize}
The cross section result as a function of $E_T^c$ is shown in \Fig{res-trilep}. Heavy resonance searches in the trilepton channel suffer from the small branching ratio into three leptons. For 2-TeV resonance, the cross section is well below $\sim 0.1$fb (which gives 10 number of signal events) in most of parameter space. For case B and C, $M_{KK} \simeq 1600-1700 \GeV$ is within reach with a mild $E_T^c \simeq 150-200 \GeV$ cut. However, $E_T^c \simeq 200-250 \GeV$ can probe strong coupling case A still up to about $M_{KK} \simeq 1900 \GeV$.

Second, we consider more relaxed event selections following an ATLAS report (see page 1603 of ref. \cite{Aad:2009wy}).
\begin{itemize}
 \item Tri-lepton event selection set $\#2$:
  \begin{enumerate}
    \item \underline{At least} 3 leptons with $p_T \geq 10 \GeV, \, |\eta|\leq 2.5, \, \Delta R_{lj} \geq 0.4$
    \item \underline{At least} 1 jet with $p_T \geq 200 \GeV, \, |\eta|\leq 2.5$ , no b-tagging.
    \item \underline{No} $E_T^{miss}$ cuts
  \end{enumerate}
\end{itemize}
After cuts, $\sigma(\tT) \simeq 11\xfb$ and $\sigma(ZW) \simeq 1\xfb$ remain dominant backgrounds. The signal cross section of $1.7\xfb$ is required after cuts for $5 \sigma$ discovery with $100 \xfb^{-1}$ of data. Signal efficiencies are given in the Table \ref{res-trilep2}. $5 \sigma$ discovery is possible for $M_{KK} \lesssim 1450 \GeV$, which is  lower than what can be obtained using the set $\#1$ cuts above.

\begin{table}[t] \centering \begin{tabular}{|c|c|c|c|c|c|}
\hline
$M_{KK}$  & 1000 GeV & 1500 GeV & 2000 GeV & 2500 GeV & 3000 GeV \\
\hline
efficiency of cut $\#2$ & 1.66 $\%$ & 1.11 $\%$ & 0.75 $\%$ & 0.65 $\%$ & 0.47 $\%$ \\
\hline
$\sigma(\tT \tT)_A \cdot$ eff & 36 fb & 1.2 fb & 0.08 fb & 0.01 fb & 0.001 fb \\
\hline
\end{tabular} \caption{Tri-lepton results with trilepton cut $\#2$. Case A signal cross section is shown.} \label{res-trilep2} \end{table}

ref. \cite{Aad:2009wy} (pages 1604-1605) has studied another set of cuts using $E_T^{miss}$ but without any jet requirements. This might be more efficient for some supersymmetry scenarios. However, a large background of $\sim 70\xfb$ remains. Due to the small four-top cross sections, this approach is not suitable for us.

\subsection{Identification of boosted leptonic top quark}
\label{sec-leptop}

Some leptons come from a top quark which is a daughter of $\KKg$. This $\KKg$'s daughter top quark is likely to be boosted with high $p_T \sim M_{KK}/2 \sim 1 \TeV$, and its decay products are likely to be collimated. In the detector, a lepton is typically well-defined when it is well isolated from jetty activities (e.g., see ref.\cite{Aad:2009wy}). So many leptons might be lost. In this subsection, we study how serious the lepton-jet collimation issue is in our multi-lepton observables, and discuss possible improvements based on leptonic top tagging. Since the single lepton channel has a quantitatively different answer, we discuss them separately at the end of the subsection.

First, we compute how much leptons overlap with some jets in the LSDL and trilepton final states. Table \ref{tbl-lepiso} shows how event selection efficiency changes when lepton isolation criteria is loosened $\Delta R_{lj} \geq 0.4 \to 0.2$ (with event selection $\#1$'s). We see that in many LSDL events and the majority of trilepton events there are non-isolated leptons with the standard isolation criteria. Moreover, collimation becomes more important for a heavier resonance as its daughter top will be more boosted.

\begin{table}[t] \centering \begin{tabular}{|c|c|c|}
\hline
LSDL & $E_T^c = 100 \GeV$ & 300 GeV \\
\hline
$M_{KK} = 1500 \GeV$ & $2.3 \% \to 3.7 \%$ & $0.75 \% \to 1.3 \%$ \\
$M_{KK} = 2000 \GeV$ & $2.0 \% \to 3.7 \%$ & $0.85 \% \to 1.7 \%$ \\
$M_{KK} = 2500 \GeV$ & $1.7 \% \to 3.5 \%$ & $0.79 \% \to 1.8 \%$ \\
\hline \hline
trilepton & $E_T^c = 100 \GeV$ & 300 GeV \\
\hline
$M_{KK} = 1500 \GeV$ & $0.81 \% \to 1.8 \%$ & $0.21 \% \to 0.59 \%$ \\
$M_{KK} = 2000 \GeV$ & $0.51 \% \to 1.7 \%$ & $0.15 \% \to 0.72 \%$ \\
$M_{KK} = 2500 \GeV$ & $0.52 \% \to 1.8 \%$ & $0.19 \% \to 0.89 \%$ \\
\hline
\end{tabular} \caption{Efficiency changes by $\Delta R_{lj} \geq 0.4 \to 0.2$ to see lepton-jet collimation. LSDL, trilepton event selection $\#1$'s are used, respectively.} \label{tbl-lepiso} \end{table}

Second, we estimate how much lepton-jet collimation is really due to boosted top quark decays (not by random overlapping). If it is, lepton will be collimated with the  $b$-jet from the same top quark (without parton showering). Table \ref{tbl-toplep} shows the change of efficiency when such a lepton is also counted as an isolated lepton. If a boosted leptonic top can be efficiently identified, we can count such a leptonic top as an isolated lepton. ref.~\cite{Rehermann:2010vq} has recently discussed such efficient leptonic top tagging where tagging efficiency of $\sim 80\%$ is obtained with a rejection $\sim 10^3 - 10^4$.

Results with standard isolation $\Delta R_{lj} \geq 0.4$ is shown in Table \ref{tbl-toplep} to see the importance of the collimation problem in the future measurements based on such standard isolation criteria. Efficient id of leptonic top can then enhance the trilepton signal events by a large factor of $2.5-6$, and the LSDL signal by about $1.5-2.5$ with standard $\Delta R_{lj} \geq 0.4$. Leptonic top id will be more useful for heavier $\KKg$ with higher $E_T^c$ as can be seen by the higher rate of increase in the table. It is clearly because collimation is due to high $p_T$ boosted objects; the heavier $\KKg$, the more boosted top, and high $E_T^c$ makes us focus more on such high $p_T$ objects. In addition, signal-to-background ratio is smaller with higher $E_T^c$ (see \Fig{res-lsdl} and \Fig{res-trilep}). By comparing $\Delta R_{lj} \geq 0.3$ values in Table \ref{tbl-toplep} with \Fig{res-lsdl} and \ref{res-trilep} (recall that event selection $\#1$'s use this lepton isolation), we estimate that $5 \sigma$ reach can be extended by about $100-200 \GeV$. By comparing Tables~\ref{tbl-lepiso} and \ref{tbl-toplep}, we also see that most lepton-jet collimation happens inside a leptonic top jet. Given this potential improvements, it would be interesting to do more detailed study of id of boosted leptonic objects.

\begin{table}[t] \centering \begin{tabular}{|c|c|c|}
\hline
LSDL & $E_T^c = 100 \GeV$ & 300 GeV \\
\hline
$M_{KK}=1500 \GeV$ & $2.3\%(3.1) \to 3.5\%$ & $0.75\%(1.0) \to 1.2\%$ \\
$M_{KK}=2000 \GeV$ & $2.0\%(2.9) \to 3.9\%$ & $0.85\%(1.3) \to 1.9\%$ \\
$M_{KK}=2500 \GeV$ & $1.7\%(2.4) \to 3.8\%$ & $0.79\%(1.3) \to 2.1\%$ \\
\hline \hline
trilepton & $E_T^c = 100 \GeV$ & 300 GeV \\
\hline
$M_{KK}=1500 \GeV$ & $0.81\%(1.3) \to 1.8\%$ & $0.21\%(0.39) \to 0.65\%$ \\
$M_{KK}=2000 \GeV$ & $0.51\%(1.0) \to 1.9\%$ & $0.15\%(0.41) \to 0.93\%$ \\
$M_{KK}=2500 \GeV$ & $0.52\%(1.1) \to 2.2\%$ & $0.19\%(0.48) \to 1.1\%$ \\
\hline
\end{tabular} \caption{Efficiency changes by leptonic top quark id (refer to text) to see lepton-$b$ collimation inside a top jet. $\Delta R_{lj} \geq 0.4 (0.3)$ with event selection $\#1$'s are used. Final efficiency is almost the same for both lepton isolations.} \label{tbl-toplep} \end{table}

The single lepton observable, on the other hand, would not take advantage of leptonic top tagging. From Table~\ref{tbl-sincoll}, we see that single lepton event samples become rather smaller with the looser lepton isolation condition $\Delta R_{lj} \geq 0.4 \to 0.2$, or with efficient leptonic top tagging shown in the last column (in the last column, as we did before, we include lepton if it is collimated with the $b$-jet from the same top quark). This is because many single lepton events are actually contributed from multi-lepton events ($n_l \geq 2$) by losing some of their leptons. So including more leptons would take these contributions out of the single lepton sample pool. Given this observation, we try to tighten the lepton isolation criteria $\Delta R_{lj} \geq 0.4 \to 0.6$ to see if we can have larger single lepton samples. The answer is no as shown in Table~\ref{tbl-sincoll}. Thus, we find that the standard lepton isolation criteria is suited for the single lepton observable that we employ.

\begin{table}[t] \centering
  \begin{tabular}{|c|c|c|c|}
    \hline
    Single lepton $M_{KK} = 2\TeV$ & $\Delta R_{lj} \geq 0.4 \to 0.2$ & $\Delta R_{lj} \geq 0.4 \to 0.6$ & lepton-b jet inclusion \\
    \hline
    $E_T^c = 200 \GeV$ & $7.04 \% \to 6.56 \%$ & $ 7.04 \% \to 6.02 \%$ & $7.04 \% \to 5.82 \%$ \\
    \hline
    $E_T^c = 400 \GeV$ & $2.43 \% \to 2.39 \%$ & $ 2.43 \% \to 2.01 \%$ & $2.43 \% \to 2.04 \%$ \\
    \hline
  \end{tabular}
\caption{ Efficiency changes due to several modifications described in the first row for two choices of $E_T^c$. Single lepton event selection $\#1$ is used with $M_{KK} = 2000 \GeV$ to produce this table.} \label{tbl-sincoll}\end{table}

\section{When is quadruple top production most important?}

We discuss other possible collider signatures in the present model. One type of signature is based on the prospect  that $\KKg$ couplings to light quarks might be somehow induced. Also, there are many other KK particles that can be very light or can interact with light fermions. Our goal is to describe the parameter space where the four top production is the most important.

\subsection{Light quark couplings to $\KKg$}

The standard search channel $q\bar{q} \to \KKg \to t\bar{t}$ at the LHC will be dominant over four top production even for small light quark couplings. How close should $c_{light}$ be to $0.5$ to suppress the standard channel enough? It is usually claimed that $M_{KK} \lesssim 4-5 \TeV$ with $g_{light}^{old} \simeq 0.2 g_{QCD}$ can be accessible
around the resonance $m_{\tT} \simeq M_{KK} \pm \Gamma_{KK}$ \cite{Lillie:2007yh}. The $\tT$ production ratio of a 2-TeV and 5-TeV $\KKg$ is approximately $\sigma_{S2} / \sigma_{S5} \, \simeq \, 100-200$ with similar signal to background ratios $S_2 / B_2 \, \sim \, S_5 / B_5$ after signal kinematic cuts (with hard jet cuts or mildly efficient top-tagging rejecting QCD dijet by a factor of $\sim 10$) \cite{Lillie:2007yh}. We find the required suppression factor $\epsilon = ( g_{light}^{our} / g_{light}^{old} )^2$ ($g_{light}^{our}$ is our smaller couplings) of the 2-TeV cross section to have statistical significance similar to that of 5-TeV signal is
\beq
\frac{ (\epsilon \sigma_{S2})\, {\cal L} }{ \sqrt{\sigma_{B2} {\cal L}} }  \, \simeq \, \frac{ \sigma_{S5} {\cal L} }{ \sqrt{\sigma_{B5} {\cal L}} }.
\eeq
Using signal to background ratios quoted above, we obtain
\beq
\epsilon \sqrt{ \sigma_{S2} {\cal L} } \, \simeq \, \sqrt{ \sigma_{S5} {\cal L} }
\eeq
which gives
\beq
\epsilon \, \simeq \, \sqrt{ \frac{ \sigma_{S5}}{\sigma_{S2}} } \, \sim \, \frac{1}{10} - \frac{1}{20}.
\eeq
Thus $g_{light}^{our} \lesssim 0.04 - 0.06 g_{QCD}$ will suppress the 2-TeV $\KKg$ signal in the $\tT$ channel below the discovery reach. Thus, from \Fig{coup-f0f0g1}, $0.49 \lesssim c_{light} \lesssim 0.51$ is the region where the four-top production is the primary (at least useful complimentary) channel of the RS discovery, which is our region of interest.

Another source of light quark couplings is through the mixing of gauge eigenstates. CKM matrix elements between the third and the first two generations are nonzero. From the above estimation of the range $g_{light} \lesssim 0.04 - 0.06 g_{QCD}$ with typical coupling strength of third generation $g_{top} \sim {\cal O}(1) g_{QCD}$, the mixture of third generation in the first generation should be of $\lesssim {\cal O}(1) \%$ which is fine with small CKM element $V_{td} \sim {\cal O}(0.001)$ although precise numbers might be model dependent.

Higher order corrections exist. The effective interaction vertex of $g-g-\KKg$ can be induced by strongly coupled top quark loop. As the theory is chiral, an anomaly cancellation mechanism should be specified to estimate the finite triangle loop contribution. This has been estimated to be negligible with Chern-Simons term \cite{Djouadi:2007eg}. Another effect of loop corrections to bulk masses of a few percent may exist \cite{Cheng:2002iz}. This correction may be quite small, but if it is larger than the characteristic range of $c$ that we need for four top production, it merely shifts the original value of $c_{light}$ such that after corrections the tuned value is near $c_{light}=0.5$.

\subsection{Targeting other KK particles} \label{sec:other-sig}

Successful custodial protection of the $T$ parameter and $Zb_L \bar{b}_L$ coupling is based on the custodial symmetries $SU(2)_R \times P_{LR}$. Extra KK gauge bosons therefore exist. Also, SM particles should be embedded in a full representation of the custodial symmetries. This implies that there are exotic fermions (custodians) as well. We review collider searches of such KK particles and find the region of parameter space in which four top production is primarily important.

In order to protect the $SU(2)_L$ coupling part of the bottom quark (which has been measured with most precision among third generation couplings), the SM doublet should be a bi-doublet under $SU(2)_L \times SU(2)_R$ \cite{Agashe:2006at}
\beq
\bmat t_L \\ b_L \emat \quad \To \quad \bmat t_L \,(+,+) & T_L \,(-,+) \\ b_L \,(+,+) & B_L \,(-,+) \emat
\eeq
where $SU(2)_{L(R)}$ acts vertically (horizontally). Orbifold boundary conditions are chosen in such a way that the $SU(2)_R$ is conserved on the IR brane and only SM particles have zero modes. On the other hand, the RH top should be embedded into a singlet or triplet under $SU(2)_R$ in order to have a Yukawa coupling (Higgs is residing in $(2,2)$ representation) \cite{Agashe:2006at}. If $t_R$ is embedded into a triplet, another triplet is required by $P_{LR}$. We will simply assume a singlet $t_R$.
\beq
t_R \quad \To \quad (t_R (+,+)).
\eeq
Effects of the new particles in triplets (if $t_R$ were in a triplet) on EWPT is not that significant~\cite{Carena:2006bn}.

Collider searches of exotic fermions have been carried out in many places. Electroweak singlet $t_R^1$ mixing with the top quark can be probed in its pair production followed by subsequent decays to $bW$. $5\sigma$ discovery reach is estimated to be about $M_{t_R^1} \simeq 1\TeV$ \cite{AguilarSaavedra:2005pv}.
$b_L^1$ and $T$ can also be pair produced and decay to $tW$ via mixing of their $SU(2)_L$ partners with top quark. Strategy based on jet mass can achieve $M_{b_L^1} \simeq 1 \TeV$ $5\sigma$ discovery reach~\cite{Skiba:2007fw} while usage of the LSDL observable can raise the potential $M_{b_L^1,\,T} \simeq 1.2 \TeV$ \cite{Contino:2008hi} which can also be augmented by combining single production of exotic fermions \cite{Mrazek:2009yu}.

\begin{figure}
\centering
\includegraphics[angle=0,width=0.48\textwidth]{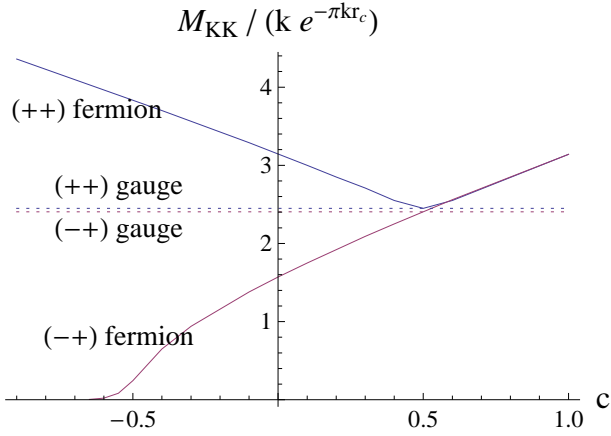}
\caption{Masses of the first KK gauge boson and (LH) fermions  in units of $M_{IR} = k e^{-\pi k r_c}$. Two types of orbifold boundary conditions $(+,+)$ and $(-,+)$ are shown. KK gauge masses are $M_{KK} \simeq 2.45(++),\, 2.405(-+) M_{IR}$. EWSB mixing effect is ignored.}
\label{mass_kk} \end{figure}

The KK spectrum of KK fermion is shown in \Fig{mass_kk} as a function of bulk mass by ignoring KK-zero mode mixing. Singlet $t_R^1$ will not be light enough for the entire range of $c_t$ considered in \Eq{eq:c-range} with $M_{KK} \gtrsim 1.5 \TeV$ because $t_R^1 \,(+,+)$ is always heavier than $\KKg$. On the other hand, for small $c_Q$, there will be very light $(-,+)$ fermions such as $T$ and $B$. For 2-TeV $\KKg$ (1.5-TeV $\KKg$), $T$ is heavy enough if $c_Q \gtrsim 0 \, (0.25)$. In all, our four top production with cases A,B and D becomes a favored discovery channel for $M_{KK} \gtrsim 1.5 \TeV$, while for case C four tops will be most important for a slightly heavier $M_{KK} \gtrsim 1.7$.

We comment that if $t_R$ is a part of a triplet $\left( \, T_R(-,+),\, t_R(+,+),\, B_R(-,+) \, \right)$ (and its $P_{LR}$ partner triplet with all fermions satisfying $(-,+)$ BC), $c_t$ will have similar preferred range as $c_Q$ above. This is because the expected discovery potentials of electroweak singlet and doublet fermions are similar. For all cases A-D with $M_{KK} \gtrsim 1.7 \TeV$, our four top production again will be a promising channel.

\begin{figure}
\centering
\includegraphics[angle=0,width=0.46\textwidth]{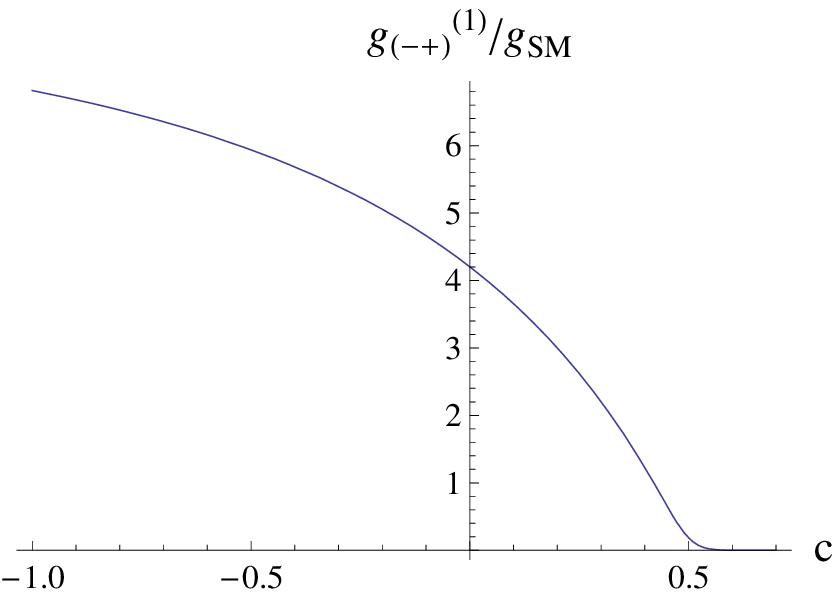}
\includegraphics[angle=0,width=0.46\textwidth]{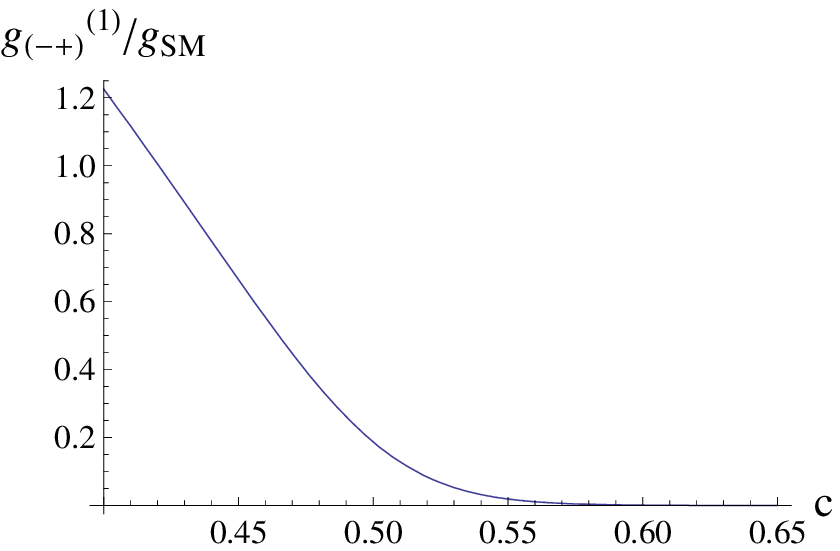}
\caption{Gauge coupling of zero mode fermions with the lowest KK gauge boson with $(-,+)$ boundary condition. $g_{KK}\simeq 0.19 \, (0.02)$ for $c=0.5 \, (0.6)$, and approaches zero with higher $c$.}
\label{coup-f0f0g1-mp} \end{figure}

There are also KK excited gauge bosons. Gauge symmetry of a model is given by
\beq
G_{bulk} \= SU(3)_c \times SU(2)_L \times SU(2)_R \times U(1)_{B-L} \, \to \, G_{SM} \= SU(3)_c \times SU(2)_L \times U(1)_Y
\eeq
where symmetry breaking $SU(2)_R \times U(1)_X \to U(1)_Y$ is through orbifold boundary condition $(-,+)$. Gauge bosons associated to broken parts do not have zero modes and the lightest KK modes are denoted by $\widetilde{W}^{1,2}, Z^\prime$ \cite{Agashe:2003zs}. Due to this boundary condition, $\widetilde{W}, Z^\prime$ are not orthogonal to flat wave functions. Consequently, they couple to light fermions even for $c_{light}=0.5$. See \Fig{coup-f0f0g1-mp} for their couplings to zero mode fermions: its coupling strength is $g_{KK} \sim 0.2 g_{weak}$ for $c=0.5$, where $g_{weak}$ is a SM weak gauge coupling. Masses of $\widetilde{W}$ and $Z^\prime$ are shown in \Fig{mass_kk}. We ignore any bulk breaking of these symmetries so that masses are determined by boundary conditions $(-,+)$. Additional bulk breaking may raise masses of these KK gauge bosons, and make collider search of these KK bosons unavailable.

$q\bar{q} \to Z^\prime \to \tT$ is smaller than the usual KK gluon-mediated $\tT$ production by a factor of $\sim (1/6)^2 \cdot 0.5 \sim 0.01-0.02$. $1/6$ is the ratio of weak gauge and QCD couplings, and $Br(Z^\prime \to \tT) \sim 0.45$ which is about half the usual KK gluon case giving additional suppression factor of $0.5$ if $c_Q \ne 0.5$. The branching ratio is reduced because the bottom quark coupling same size of top coupling is turned on if $c_Q \ne 0.5$. Then the 2-TeV $Z^\prime$ mediated $\tT$ cross section is similar in magnitude as the usual 5-TeV KK-gluon mediated one. Given that $m_{\tT}$ around 2 TeV will be submerged into a larger QCD background than the 5 TeV case, we conclude that $Z^\prime$ mediated $\tT$ production may not be a promising channel. $\widetilde{W}$ may contribute to single top production as the $W$ boson does in the SM. For instance, the process $u\bar{d} \to \widetilde{W} \to t\bar{b}$ is smaller than SM $W$-mediated process by $\sim (4 \cdot 0.2)^2 \cdot M_W^2/M_{KK}^2 \sim 0.002$ where $g_t/g_{weak} \sim 4$ and $g_u/g_{weak} \sim 0.2$. So some powerful discriminator is needed. Given this difficulty and the possibility of raising the mass of $\widetilde{W}$ by bulk breaking, we conclude that we have higher sensitivity through the four-top production process.

\section{Conclusion}

We have studied four-top signatures of Randall-Sundrum model in the case of $c_{light} \simeq 0.5$ with universal RH down sector $c_b = c_{light}$. Associate production of $\KKg$ with $\tT$ as well as pair production of $\KKg$ can produce four top quarks. We have estimated the discovery reach in the single-lepton, like-sign dilepton, trilepton final states of four-top events. For a strongly coupled right-handed top case, the like-sign dilepton observable has the highest potential that can probe up to $M_{KK} \sim 2-2.5 \TeV$. On the other hand, for a strongly coupled left-handed top case, the single-lepton observable,  which is enhanced by $\tT b\bar{b}$ events via $\KKg$ associated production, is the most promising channel for $M_{KK} \lesssim 2 \TeV$.

In the LSDL and trilepton channels, boosted top and its collimated lepton-jet issue arise. Efficient identification of boosted leptonic top quark can enhance the number of signal events by a factor of about $2$(LSDL) and $4$(trilepton) with standard isolation $\Delta R_{lj} \geq 0.4$ as illustrated in Table \ref{tbl-toplep}. This will be more effective with higher $E_T^c$ cuts. A more detailed study of leptonic top id is well motivated. On the other hand, the implications of lepton-jet collimation is different in the case of single lepton final state. Since many multi-lepton events contribute to single lepton event samples by losing some of their leptons, efficient id of leptonic objects can rather degrade the single lepton sample pool (to the benefit of other channels).

The strongly coupled LH case considered in this paper (coupling sets C and D) can represent the favored parameter space found in the previous literature when considering EWPT constraints and also the flavor-shining model of ref.\cite{Delaunay:2010dw}. Due to this importance and the relatively large signal cross sections, we have also studied an alternative single-lepton observable composed of three $b$-quark tags. Although detailed background estimation by experiment is required,  we have estimated the  discovery potential to be up to  $\sim 2.4-2.8 \TeV$ with assumed $b$-tagging efficiency $\epsilon_b = 0.4 -0.6$.

We have also discussed competing signatures from custodians and KK gauge bosons of custodial symmetry $\widetilde{W}, Z^\prime$. Unless $c_{t,Q} \lesssim 0$, custodians are not light enough and their pair productions are small. $Z^\prime, \widetilde{W}$ mediated top production is suppressed by their weak gauge coupling nature. In large parameter space near $c_{light}=0.5$ our four-top signal dominates.

$c_{light}=c_b=0.5$ more or less gives up the geometric approach to flavor physics in the collider-reachable sectors in the warped model. However, increasing tension with precision data and the ensuing tensions of a fine-tuned weak scale  make deserving the study of $c_{light}=0.5$, where many phenomenological issues are relieved. As we have discussed, this approach may significantly reduce the ability to find KK gluons through resonance production from light quarks, and four top quark events may in the end be the best path to discovery.
Said a different direction, if a very light KK gluon $M_{KK} \sim 1.5-2 \TeV$ is realized, $c_{light}=0.5$ is likely to be Nature's choice and four-top production via KK-gluons may be the first beyond the SM discovery signature.

\section*{Acknowledgement}
We appreciate discussions with K. Agashe, T. Gherghetta, V. Sanz, J. Serra, G. Servant, M. Toharia and A. Weiler. SJ is supported in part by a Samsung Corporation scholarship and by the U.S. Department of Energy.


\end{document}